\documentclass[12pt]{article}
\pdfoutput=1
\usepackage{tikz}
\usepackage{a4wide,epsfig,psfrag,amsmath,amssymb,cite,scalefnt}
\usepackage{color}
\usepackage{amsmath,comment,braket}
\usepackage{placeins}

\graphicspath{ {figs/} }

\newcommand{\ep}{\epsilon}

\newcommand{\evenspace}{ \vphantom{\Bigg\{} }
\newcommand{\api}{a_s}
\newcommand{\GF}{G_F}
\newcommand{\Mh}{m_H}
\newcommand{\Mt}{M_t}
\newcommand{\nh}{n_h}
\newcommand{\nl}{n_l}
\newcommand{\rr}{\rho}
\newcommand{\cAC}{C_A}
\newcommand{\cRC}{C_F}
\newcommand{\ItRC}{T_f}
\newcommand{\lmuMh}{L_{\Mh^2}}
\newcommand{\lmuMt}{L_{\Mt^2}}
\newcommand{\lmumuf}{L_{\mu_f^2}}

\allowdisplaybreaks

\newcommand{\gsim}{\;\rlap{\lower 3.5 pt \hbox{$\mathchar \sim$}} \raise 1pt
 \hbox {$>$}\;}
\newcommand{\lsim}{\;\rlap{\lower 3.5 pt \hbox{$\mathchar \sim$}} \raise 1pt
 \hbox {$<$}\;}

\begin{document}

\title{\vskip-3cm{\baselineskip14pt
    \begin{flushleft}
      \normalsize TTP19-010\\
      \normalsize P3H-19-009
  \end{flushleft}}
  \vskip1.5cm
  Real-virtual corrections to Higgs boson pair production
  at NNLO: three closed top quark loops
}

\author{
  Joshua Davies$^{a}$,
  Florian Herren$^{a}$,
  Go Mishima$^{a,b}$,
  Matthias Steinhauser$^{a}$
  \\[1mm]
  {\small\it $^a$Institut f{\"u}r Theoretische Teilchenphysik}\\
  {\small\it Karlsruhe Institute of Technology (KIT)}\\
  {\small\it Wolfgang-Gaede Stra\ss{}e 1, 76128 Karlsruhe, Germany}
  \\[1mm]
  {\small\it $^b$Institut f{\"u}r Kernphysik}\\
  {\small\it Karlsruhe Institute of Technology (KIT)}
  \\
  {\small\it Hermann-von-Helmholtz-Platz 1, 76344 Eggenstein-Leopoldshafen, Germany}
}
  
\date{}

\maketitle

\thispagestyle{empty}

\begin{abstract}

  We compute the real-radiation corrections to Higgs boson pair production at
  next-to-next-to-leading order in QCD, in an expansion for large top quark mass. We
  concentrate on the radiative corrections to the interference contribution
  from the next-to-leading order one-particle reducible and the leading order
  amplitudes.  This is a well defined and gauge invariant subset of the full
  real-virtual corrections to the inclusive cross section.  We obtain analytic
  results for all phase-space master integrals both as an expansion around the
  threshold and in an exact manner in terms of Goncharov polylogarithms.
  
%

\end{abstract}

\thispagestyle{empty}

\sloppy


\newpage


\section{Introduction}

Within the Standard Model (SM) of particle physics, knowledge of the Higgs
boson mass ($m_H$) fixes the parameters of the scalar potential, since the
trilinear and quartic couplings can be expressed through $m_H$ and the vacuum
expectation value of the scalar doublet. However, in many extensions of the
SM, the trilinear and quartic couplings deviate significantly from the SM
values.  One way of probing the Higgs self coupling is through the production
of Higgs boson pairs at the LHC.  Besides the efforts undertaken on the
experimental side, also precise predictions from the theoretical side are
necessary to scrutinize the results of upcoming measurements. The most
important production mode of a pair of Higgs bosons at hadron colliders is by
top quark--mediated gluon fusion. In this channel the QCD corrections are large and
higher-order computations are important.

The leading order (LO) cross section has been known with full top quark--mass
dependence for more than 30 years~\cite{Glover:1987nx,Plehn:1996wb}.  Exact
next-to-leading order calculations are numerically quite challenging and have
only become available fairly
recently~\cite{Borowka:2016ehy,Borowka:2016ypz,Baglio:2018lrj}.
Analytic NLO calculations have so far only been performed for various
approximations. Among them is the effective theory calculation in which the heavy
top quark is integrated out~\cite{Dawson:1998py}. This result has been
extended in Refs.~\cite{Grigo:2013rya,Degrassi:2016vss} where inverse top
quark mass corrections have been computed. An expansion for small top quark
masses has been obtained in
Refs.~\cite{Davies:2018ood,Davies:2018qvx,Mishima:2018olh} and the limit of
small transverse momentum is covered by the results of
Ref.~\cite{Bonciani:2018omm}. In Ref.~\cite{Grober:2017uho} an approximation
method for the reconstruction of the form factors has been suggested. Results
from various kinematic regions are combined using conformal mapping and
Pad\'e approximation.  Finite top quark mass effects to the real radiation
contribution have been considered in Ref.~\cite{Maltoni:2014eza}.

At next-to-next-to-leading order (NNLO), the effective-theory calculation of
the cross section has been performed in
Refs.~\cite{deFlorian:2013uza,deFlorian:2013jea,Grigo:2014jma} and an
expansion for large top quark masses has been performed in
Ref.~\cite{Grigo:2015dia} in the soft-virtual approximation.  Beyond
the infinite top mass limit real radiations are missing so far;  it is the aim
of this paper to partly close this gap.  Note that in the effective-theory
calculation a large part of the corrections to Higgs boson pair production can
be taken over from single Higgs
production~\cite{Harlander:2002wh,Anastasiou:2002yz,Ravindran:2003um}; this is
no longer the case once one goes beyond this approximation.

Let us mention that recently two building blocks of the
next-to-next-to-next-to-leading order (N$^3$LO) effective-theory result have
become available: two-loop virtual corrections have been obtained in
Ref.~\cite{Banerjee:2018lfq} and the four-loop matching coefficient for the
effective coupling of two Higgs bosons and gluons has been computed
in~\cite{Spira:2016zna,Gerlach:2018hen}.

\begin{figure}[t]
  \begin{center}
    \begin{tabular}{cccc}
      \includegraphics[width=0.22\textwidth]{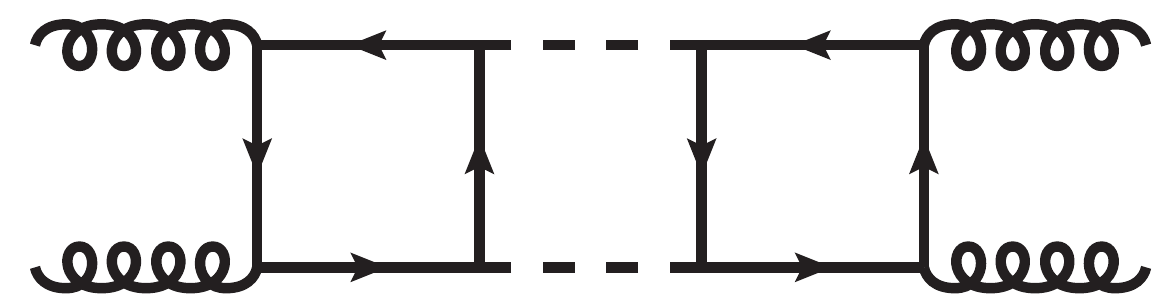}     &
      \includegraphics[width=0.22\textwidth]{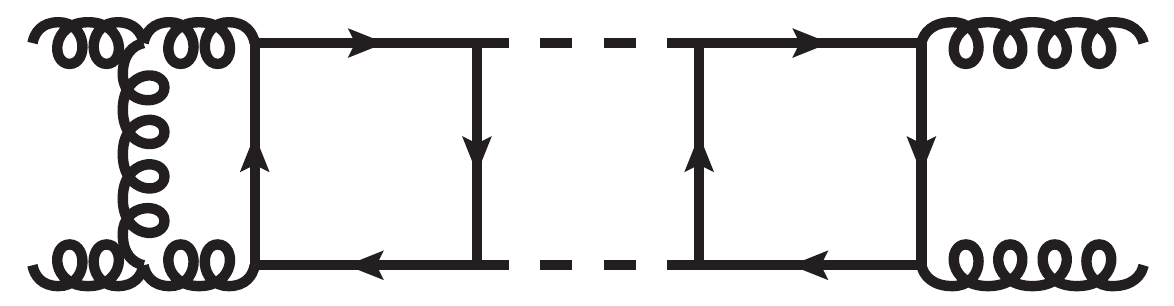} &
      \raisebox{-1em}{\includegraphics[width=0.22\textwidth]{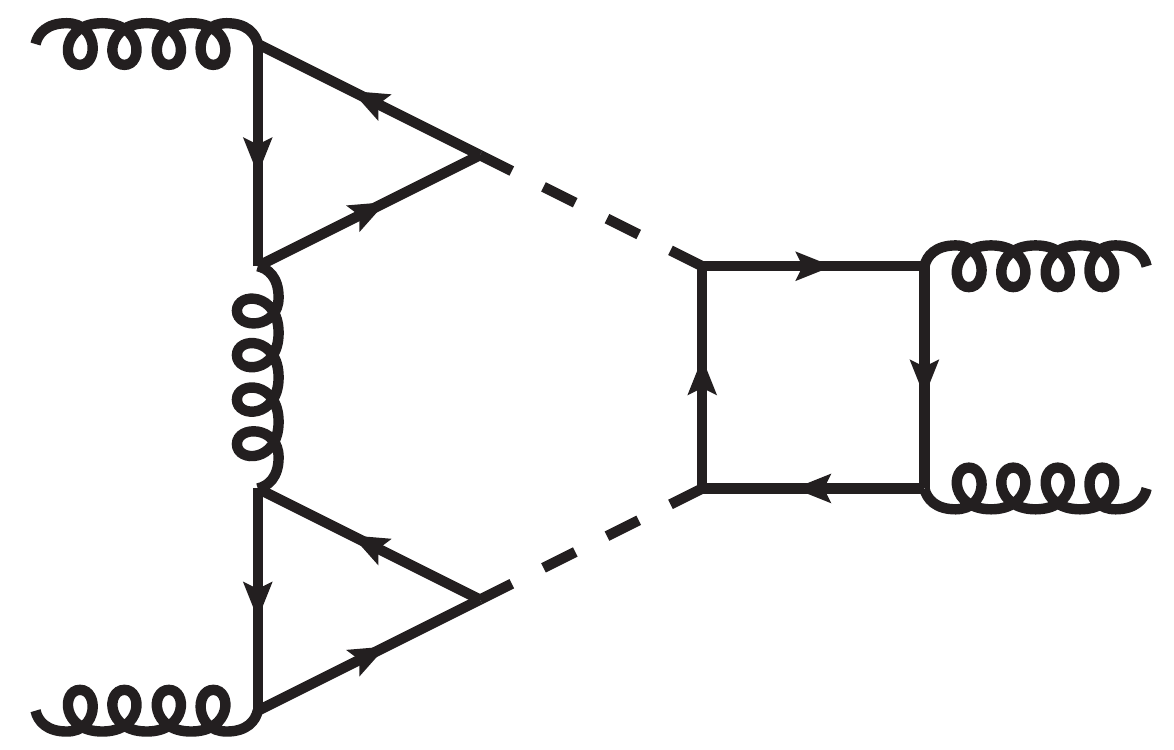}} &
      \includegraphics[width=0.22\textwidth]{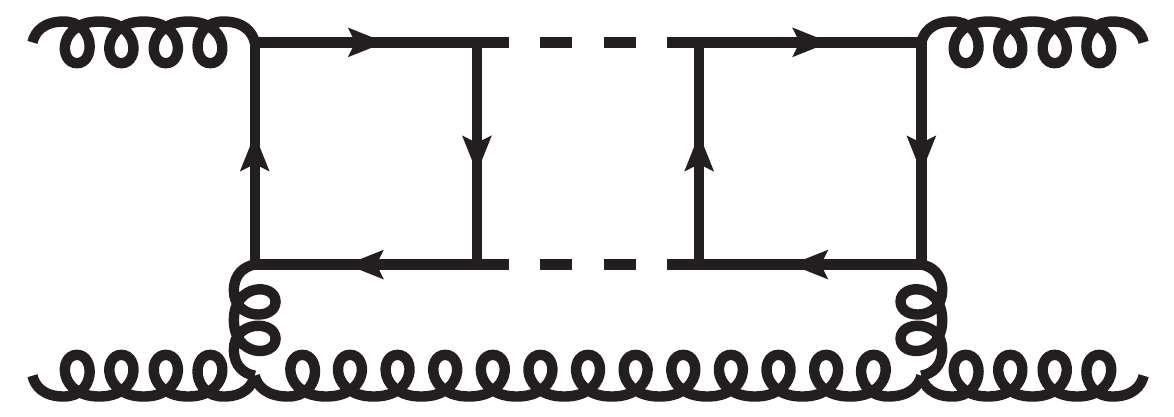} \\
      (a) & (b) & (c) & (d) \\
      \raisebox{-1em}{\includegraphics[width=0.22\textwidth]{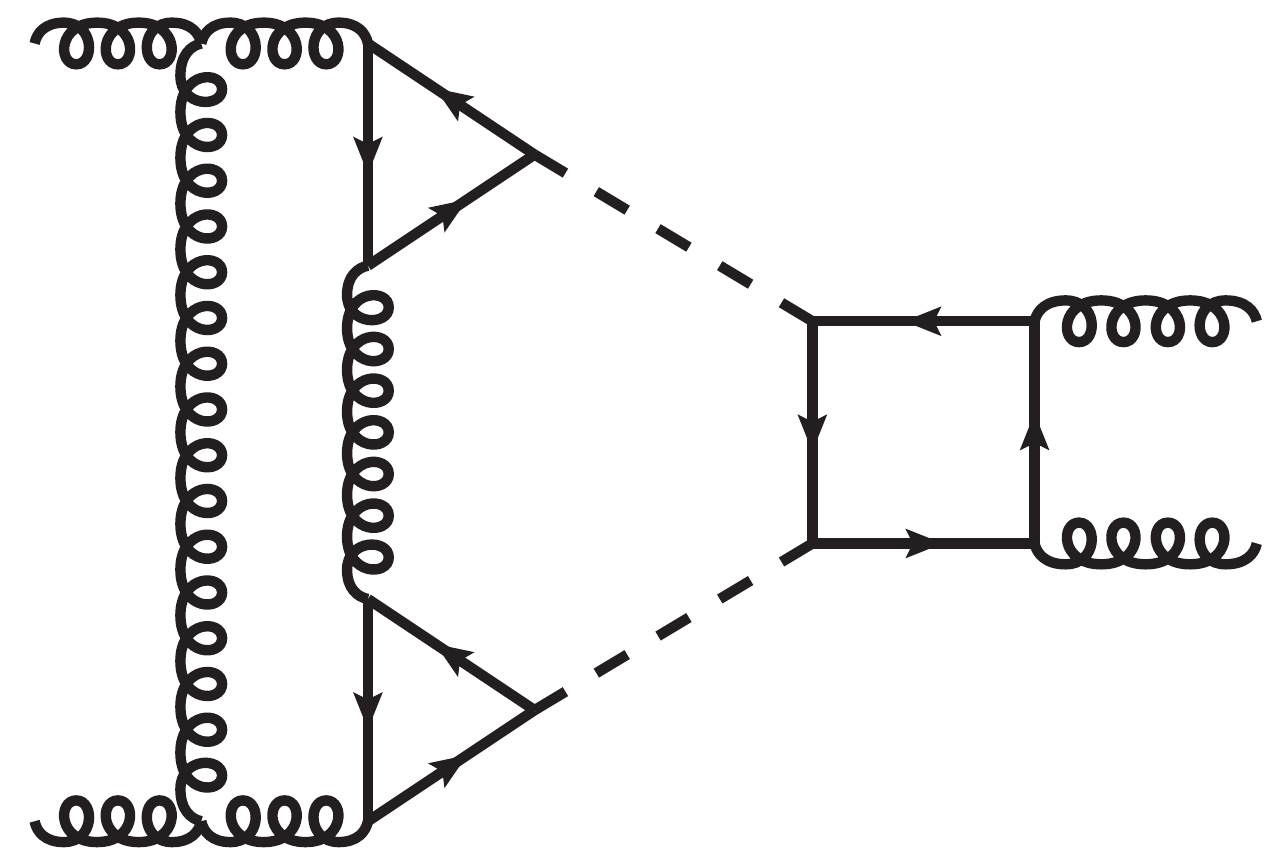}}  &
      \raisebox{-1em}{\includegraphics[width=0.22\textwidth]{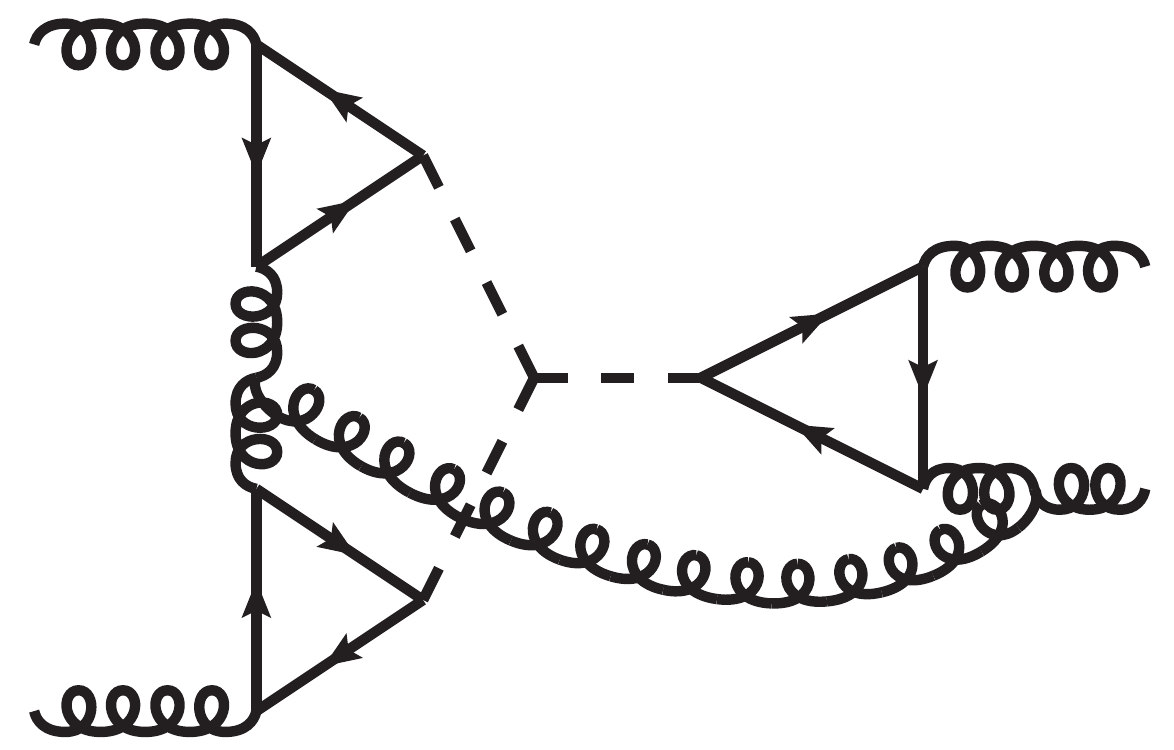}}  &
      \includegraphics[width=0.22\textwidth]{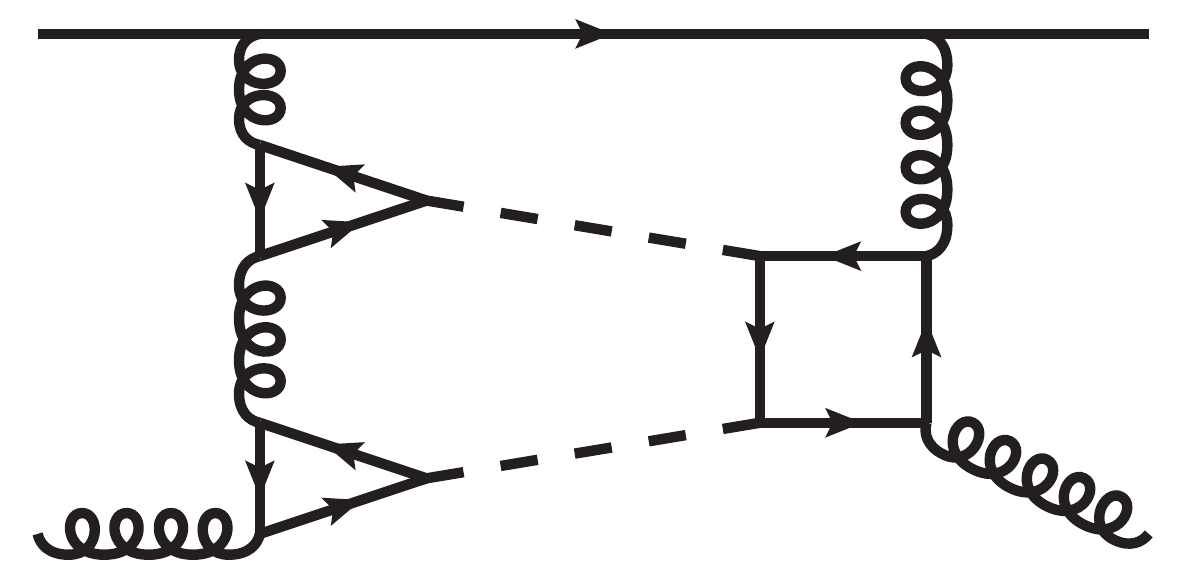} &
      \raisebox{1em}{\includegraphics[width=0.22\textwidth]{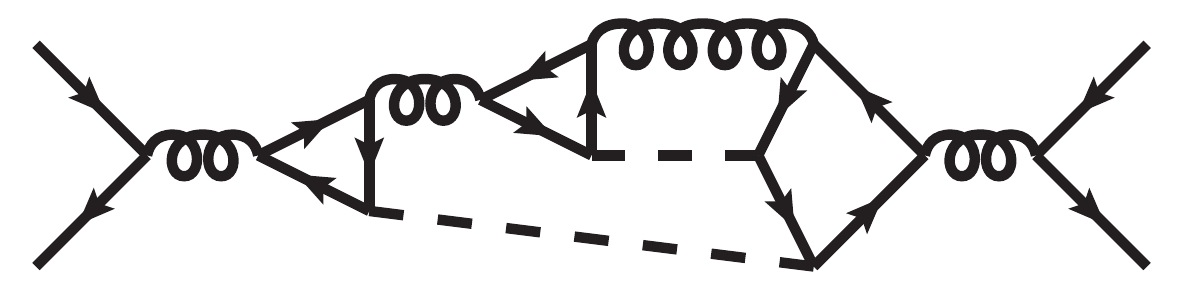}} \\
      (e) & (f) & (g) & (h) \\
      \includegraphics[width=0.22\textwidth]{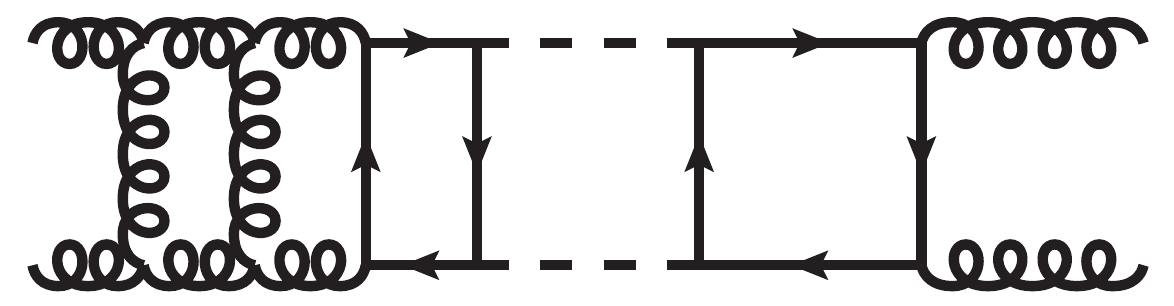} &
      \includegraphics[width=0.22\textwidth]{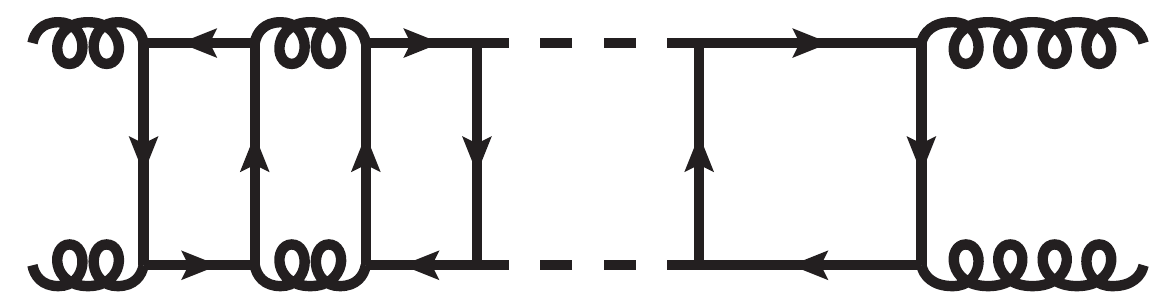} &
      \includegraphics[width=0.22\textwidth]{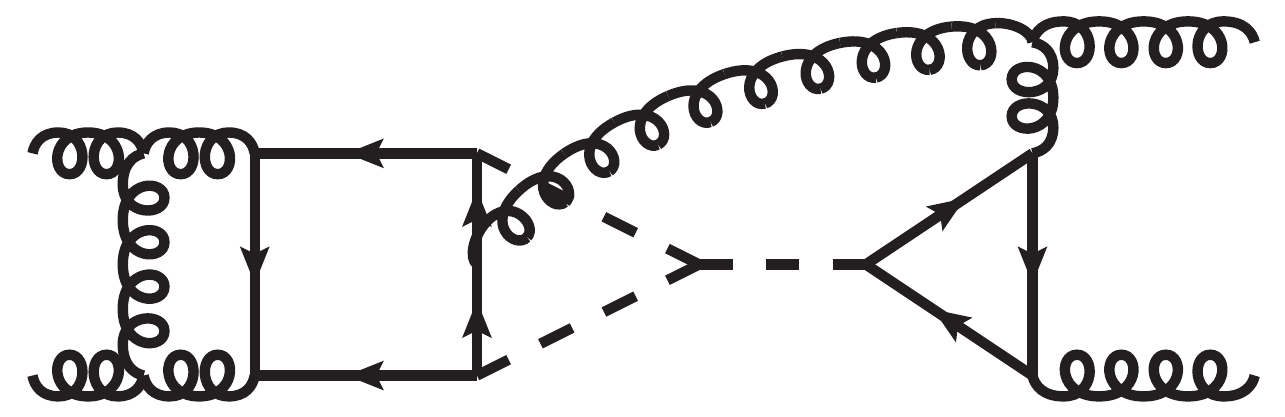} &
      \includegraphics[width=0.22\textwidth]{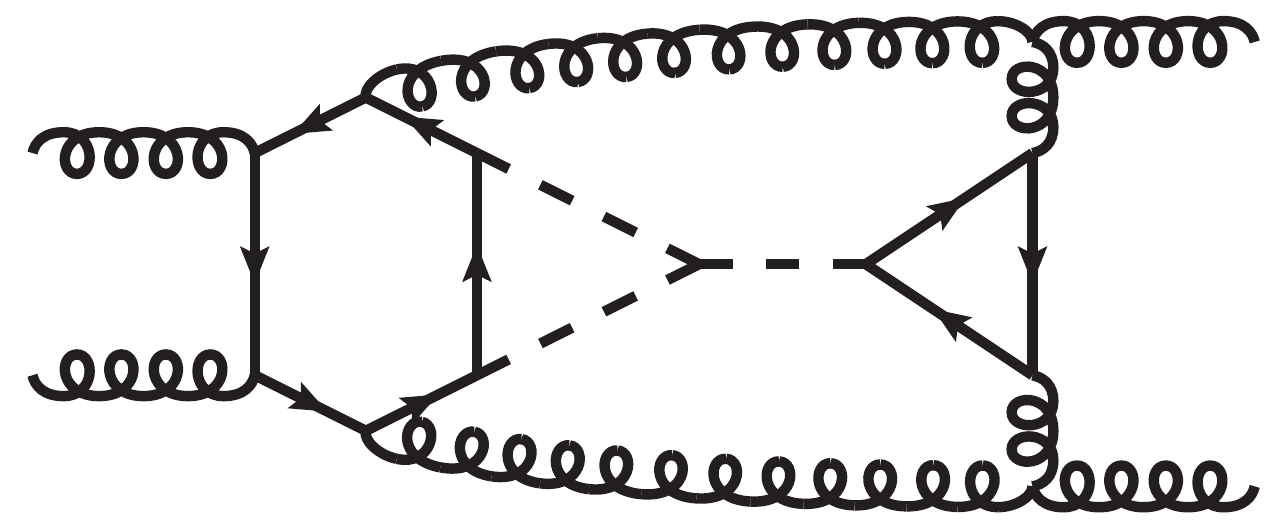} \\
      (i) & (j) & (k) & (l)
    \end{tabular}
  \end{center}
  \caption{\label{fig::sample_FDs} Sample Feynman diagrams for $ij\to
    ij$ with $i,j\in\{g,q\}$. Solid, dashed and curly lines represent
    quarks, Higgs bosons and gluons, respectively.  The first line
    contains LO and NLO contributions. NNLO contributions are shown in
    the second and third lines. The contributions to the Higgs boson
    pair production cross section is obtained by considering cuts
    which involve at least two Higgs bosons.}
\end{figure}

In this paper we compute the imaginary part of the forward scattering
amplitudes $ij\to ij$, where $i$ and $j$ stand for gluons and (anti-)quarks.  With
the help of the optical theorem one obtains the (partonic) cross section
${\rm d}\sigma/{\rm d}s$ where $\sqrt{s}$ is the center-of-mass energy of the
incoming partons.  Note that at the lowest order one already has to compute
three-loop Feynman diagrams and at N$^k$LO one has to consider $(k+3)$-loop
diagrams. Sample Feynman diagrams can be found in Fig.~\ref{fig::sample_FDs}.
The virtual corrections are obtained from the contributions where exactly two
Higgs bosons are cut.  Note that besides the virtual corrections to the NLO 1PR
diagram (Fig.~\ref{fig::sample_FDs}~(e)) also diagrams such as
Fig.~\ref{fig::sample_FDs}~(j) have three closed top quark loops.
At NLO the final state of the real radiation corrections contains two Higgs
bosons and an additional parton.  At NNLO one has either one or two additional
partons in the final state.  We refer to the former as ``real-virtual''
(Fig.~\ref{fig::sample_FDs}~(f),~(g),~(h)
and~(k)) and the latter as ``double-real'' (Fig.~\ref{fig::sample_FDs}~(l)).

The real-virtual corrections can be sub-divided according to the
number of closed top quark loops which involve a coupling to one or
two Higgs bosons. At NNLO this is either two or three, as can be
seen from the Feynman diagrams in Fig.~\ref{fig::sample_FDs}. We will
refer to them as $n_h^2$ and $n_h^3$ contributions in the following.
In this paper we consider only the $n_h^3$ contribution, with three closed top quark
loops. In an asymptotic expansion in large $M_t$ all top quark lines
are part of the so-called hard subgraphs, which means that the
remaining Feynman diagrams which involve the Higgs bosons are either
one- or two-loop diagrams.

At NLO, $n_h^3$ terms are only present in the virtual corrections, see
Fig.~\ref{fig::sample_FDs}(c). They serve as an {\it effective} LO
contribution for the $n_h^3$ NNLO corrections we are interested in. In this
sense, one can consider the subset of real-virtual corrections with three top
quark loops as {\it effective} NLO real corrections. Thus, they share many
features with the NLO real corrections and many steps of the calculation can
be performed in analogy to Ref.~\cite{Grigo:2013rya}.

We consider all partonic channels contributing to the cross section and
compute the necessary phase-space integrals both as an expansion around the
pair-production threshold~\cite{Grigo:2013rya} and also as expressions exact in $m_H$ and
$s$. In the latter case the analytic results are expressed in terms of
Goncharov polylogarithms~\cite{Goncharov:1998kja}.  As we will show, for all
phenomenological applications it is sufficient to consider the expanded
results which have a simpler mathematical structure.

While for single Higgs production top quark mass suppressed terms converge
well after factoring out the exact LO cross section (see
Refs.~\cite{Pak:2009dg,Harlander:2009my} for the NNLO analysis), this is not
the case for Higgs boson pair production~\cite{Grigo:2013rya}. However, the
large top quark mass expansion can still provide valuable input for
approximation methods in the full range of center of mass
energies~\cite{Grober:2017uho,Grazzini:2018bsd}.

The remainder of this paper is organized as follows. In the next Section we
introduce our notation and comment on the techniques which have been used for
the calculation.  We discuss our results in Section~\ref{sec:res} and present
our conclusions in Section~\ref{sec:conclusion}.  In the Appendix we provide
additional material such as details on the threshold expansion of the
phase-space master integrals (Appendix~\ref{app::MI_delta}) and the calculation
of the master integrals without expansion (Appendix~\ref{app::MI_exact}).


\section{\label{sec:tech}Preliminaries and technicalities}

In this section we fix our notation and provide technical details on the
calculation of the real-virtual $n_h^3$ contribution to the cross section for
double Higgs production.


\subsection{Cross section}

We compute the cross section by applying the optical theorem to the
forward scattering amplitudes
\begin{eqnarray}
  g(q_1)g(q_2)&\rightarrow& g(q_1)g(q_2)\,,\nonumber\\
  g(q_1)q(q_2)&\rightarrow& g(q_1)q(q_2)\,,\nonumber\\
  g(q_1)\bar{q}(q_2)&\rightarrow& g(q_1)\bar{q}(q_2)\,,\nonumber\\
  q(q_1)\bar{q}(q_2)&\rightarrow& q(q_1)\bar{q}(q_2)\,,
  \label{eq::ampl}
\end{eqnarray}
where $g$ stands for gluons and $q$ and $\bar{q}$ represent generic (light)
quarks and anti-quarks.  $q_1$ and $q_2$ are the momenta of the in- and outgoing
partons with $q_1^2 = q_2^2 = 0$. Note the at NNLO, the partonic channel which
involves different quark flavours in the initial state does not yet
contribute. Of course, we only take into account diagrams which involve cuts
through two internal Higgs boson propagators.  Sample Feynman diagrams for the
amplitudes in Eq.~(\ref{eq::ampl}) can be found in
Fig.~\ref{fig::sample_FDs}. The forward scattering
amplitudes depend on the top quark mass $M_t$, the center-of-mass energy
$\sqrt{s}$ with $s=(q_1+q_2)^2=2q_1\cdot q_2$, and the Higgs boson mass $m_H$.

In order to fix the notation and the pre-factors we now discuss
the LO and NLO cross sections in detail.
The corresponding formulae for the $n_h^3$ NNLO contributions
are obtained by straightforward replacements.

We write the perturbative expansion of the partonic cross section
for Higgs boson pair production as
\begin{eqnarray}
  \sigma_{ij \to HH + X}(s,\rho) = \delta_{ig} \delta_{jg}
  \sigma_{gg}^{(0)}(s,\rho) 
  + \frac{\alpha_s}{\pi} \sigma^{(1)}_{ij}(s,\rho)
  + \left(\frac{\alpha_s}{\pi}\right)^2 \sigma^{(2)}_{ij}(s,\rho)
  + \ldots
  \,,
\end{eqnarray}
where $\alpha_s\equiv\alpha_s^{(5)}(\mu_r)$ is the strong coupling
constant in the five-flavour theory
and $ij\in\{gg,qg,\bar{q}g,q\bar{q}\}$ denote the partonic sub-channels. 
We introduce the variable
\begin{eqnarray}
  \rho &=& \frac{m_H^2}{M_t^2}
\end{eqnarray}
to parametrize the dependence of the cross section
on the Higgs boson and top quark masses.
For later convenience we also introduce the variable
\begin{eqnarray}
  \delta 
  &=& 1 - 4x \quad \mbox{with} \quad x\,\,=\,\,\frac{m_H^2}{s}
  \,,
      \label{eq::delta}
\end{eqnarray}
which is zero at threshold. 
We renormalize the top quark mass in the on-shell
scheme. Additionally we use $\mu_r$ and $\mu_f$ for the
renormalization and factorization scales, respectively.
Note that $\mu_f$ is only present in the collinear counterterm,
where we explicitly show the dependence.

At LO, the application of the optical theorem leads to
\begin{eqnarray}
  \sigma_{gg}^{(0)} &=&
  \frac{1}{2s}\left(\frac{1}{N_A (2-2\epsilon)}\right)^2
  \widetilde{\mathrm{Im}}\left(\mathcal{M}_{gg \rightarrow gg}^{(0)} \right)
  \,,
\end{eqnarray}
where $1/(2s)$ corresponds to the flux factor and the factors $N_A=N_c^2-1$ and
$(2-2\epsilon)$ originate from averaging over gluon colours and helicities,
respectively. We use the notation $\widetilde{\mathrm{Im}}$ to indicate that we compute
the imaginary part only of cuts which involve two Higgs bosons.

Note that the factor $1/2$ for the identical Higgs bosons is contained in
$\widetilde{\mathrm{Im}}(\mathcal{M}_{gg \rightarrow gg}^{(0)})$.  In our
calculation, for the sum over the gluon polarizations we use
\begin{eqnarray}
  \sum_\lambda
  \varepsilon^{(\lambda)\star}_\mu(q_1)
  \varepsilon^{(\lambda)}_\nu(q_2)
  &=& -g_{\mu\nu}
  \,.
  \label{eq::polarization}
\end{eqnarray}
As a consequence we also have to consider amplitudes with
external ghost particles, which have to be subtracted from the
pure gluon contribution. For example, at NLO we have
\begin{eqnarray}
  \sigma_{gg}^{(1)} &=&
  \frac{1}{2s}\left(\frac{1}{N_A(2-2\epsilon)}\right)^2
  \Bigg[\widetilde{\mathrm{Im}}\left(\mathcal{M}_{gg\rightarrow gg}^{(1)}\right)
    -\widetilde{\mathrm{Im}}\left(\mathcal{M}_{gc \rightarrow gc}^{(1)}\right)
    -\widetilde{\mathrm{Im}}\left(\mathcal{M}_{cg \rightarrow cg}^{(1)}\right)
    \nonumber\\&&\mbox{} -\widetilde{\mathrm{Im}}\left(\mathcal{M}_{g\bar{c}
      \rightarrow g\bar{c}}^{(1)}\right)
    -\widetilde{\mathrm{Im}}\left(\mathcal{M}_{\bar{c}g \rightarrow
      \bar{c}g}^{(1)}\right)
    -\widetilde{\mathrm{Im}}\left(\mathcal{M}_{c\bar{c} \rightarrow
      c\bar{c}}^{(1)}\right)
    -\widetilde{\mathrm{Im}}\left(\mathcal{M}_{\bar{c} c\rightarrow
      \bar{c}c}^{(1)}\right)\Bigg]\,.
  \label{eq::sigma_real}
\end{eqnarray}
Note that all contributions with one external gluon and an external ghost
or anti-ghost field are equal.

In a similar manner we obtain the partonic cross sections for the
$qg$ (and $\bar{q}g$) and $q\bar{q}$ channels: 
\begin{eqnarray}
  \sigma_{qg}^{(1)} &=&
  \frac{1}{2s}\frac{1}{N_A(2-2\epsilon)}\frac{1}{2N_c}
  \Bigg[\widetilde{\mathrm{Im}}\left(\mathcal{M}_{qg\rightarrow
      qg}^{(2)}\right)\Bigg]\,,
  \nonumber\\
  \sigma_{q\bar{q}}^{(1)} &=&
  \frac{1}{2s}\left(\frac{1}{2N_c}\right)^2
  \Bigg[\widetilde{\mathrm{Im}}\left(\mathcal{M}_{q\bar{q}\rightarrow
      q\bar{q}}^{(2)}\right)\Bigg]\,.
  \label{eq::sigma_quarks}
\end{eqnarray}
Note that we do not need to consider ghost-quark scattering, since
this only contributes starting from N$^3$LO.

The $n_h^3$ contributions are obtained in analogy to
Eqs.~(\ref{eq::polarization}),~(\ref{eq::sigma_real})
and~(\ref{eq::sigma_quarks}) by replacing the LO part by the (virtual)
NLO corrections proportional to $n_h^3$, see
Fig.~\ref{fig::sample_FDs}~(c), which plays the role of an
{\it effective} LO contribution.  In the following we denote this part
by $\sigma^{(1),n_h^3}$. The NLO contributions in the above equations
have to be replaced by the NNLO $n_h^3$ amplitudes (see
Figs.~\ref{fig::sample_FDs}~(f), (g) and (h)), which we denote by
$\sigma^{(2),n_h^3}$.  We can thus write
\begin{eqnarray}
  \sigma_{ij\to HH + X}(s,\rho) \Big|_{n_h^3} &=& \delta_{ig} \delta_{jg}
  \sigma_{gg}^{(1),n_h^3} + \sigma_{ij}^{(2),n_h^3} + \ldots \,.
                                                  \label{eq::sig_nh3}
\end{eqnarray}
with
\begin{eqnarray}
  \sigma_{ij}^{(2),n_h^3} &=&
  \sigma_{ij,\rm virt}^{(2),n_h^3}
  + \sigma_{ij,\rm real}^{(2),n_h^3}
  + \sigma_{ij,\rm coll}^{(2),n_h^3}
  \,,
\end{eqnarray}
where the ellipses in Eq.~(\ref{eq::sig_nh3}) stand for N$^3$LO terms. 
The virtual corrections $\sigma_{ij,\rm virt}^{(2),n_h^3}$ have been computed
in Ref.~\cite{Grigo:2015dia} including terms up to $\rho^2$. Recently that
calculation has been extended up to $\rho^4$~\cite{DavSte19}.
The main
aim of this paper is the computation of the real corrections
$\sigma_{ij,\rm real}^{(2),n_h^3}$ which we discuss later in
subsection~\ref{sub::workflow}.  In the next subsection we discuss
the collinear counterterm $\sigma_{ij,\rm coll}^{(2),n_h^3}$.


\subsection{Subtracting collinear divergences}

The NNLO $n_h^3$ collinear counterterm is obtained
from the convolution of the NLO cross section $\sigma_{gg}^{(1),n_h^3}$
and the gluon or quark splitting functions, which are given by
\begin{align}
P_{i j}&=\frac{\alpha_{s}}{\pi} P_{i j}^{(0)}+\mathcal{O}\left(\alpha_{s}^{2}\right)
\,,\nonumber\\
P_{g g}^{(0)}(z)&= C_{A}\left(\left[\frac{1}{1-z}\right]_{+}-2+\frac{1}{z}+z-z^{2}\right)+\beta_{0} \delta(1-z)\,,\nonumber\\
P_{g q}^{(0)}(z)&= C_{F}\left(\frac{1}{z}-1+\frac{z}{2} \right)
\,,
\end{align}
with
\begin{align}
\beta_{0}=\frac{11}{12} C_{A}-\frac{1}{3} T_f {n_l}
\,,
\end{align}
where $n_l$ is the number of massless quarks.

In the following we concentrate on the $gg$ channel. The calculations for the
(anti) quark-induced channels proceed in a similar manner. 
The convolution integral is given by
\begin{eqnarray}
  \sigma_{gg,\rm coll}^{(2),n_h^3} &=&
  \frac{2}{\epsilon}\left(\frac{\mu_r^2}{\mu_f^2}\right)^\epsilon
  \int_{1-\delta}^1 \mathrm{d}z\, P_{gg}(z)\sigma_{gg}^{(1),n_h^3}(x/z)
      \,,
\end{eqnarray}
where the factor of 2 comes from the two external gluons.  The integral over
the delta distribution in $P^{(0)}_{gg}$ is trivial and in the parts without plus
distributions we substitute $z = 1 - \delta(1-\mu)$ and subsequently expand in
$\delta$.  The integration over $\mu$ is then straightforward.
For the contribution with the plus distribution we use the relation
\begin{align}
  \int_{1-\delta}^1\mathrm{d}z
  \left[\frac{1}{1-z}\right]_+\sigma_{gg}^{(1),n_h^3}(x/z)
   =\int_{1-\delta}^1\mathrm{d}z
  \frac{\sigma_{gg}^{(1),n_h^3}(x/z)-\sigma_{gg}^{(1),n_h^3}(x)}{1-z}
  + \sigma_{gg}^{(1),n_h^3}(x)\ln(\delta)
  \,,
\end{align}
and again expand in $\delta$ after using $z = 1 - \delta(1-\mu)$.

In order to present result for the collinear counterterm
we parametrize the NLO contribution $\sigma_{gg}^{(1),n_h^3}$ as
\begin{align}
\sigma_{gg}^{(1),n_h^3}
=\sum_{n=0}^\infty \delta ^{\frac{1}{2}+n}
\left[ c_n^{(0)} 
+\epsilon (c_n^{(1)}-c_n^{(0)} \ln \delta )
\right] 
+\mathcal{O}(\epsilon^2)
\,,
\end{align}
where $c_n^{(0)}, c_n^{(1)}$ depend on
$\alpha_s, G_F, M_t, m_H$ and $\mu_r$.  We obtain for the $gg$ channel
the following infinite series representation
\begin{align}
\sigma_{gg,\mathrm{coll}}^{(2),n_h^3}
&=\frac{\alpha_s}{\pi}
\frac{C_A}{3}
\sum_{n=0}^\infty 
\sum_{j=0}^n
\delta ^{\frac{3}{2}+n}
\frac{(n-j+3)!}{(n-j)!}
\left\{
\tilde c_j
\left(
\frac{\Delta_0}{\frac{3}{2}+n}
-
\frac{\Delta_1}{\frac{5}{2}+n}
+
\frac{\Delta_2}{\frac{7}{2}+n}
-
\frac{\Delta_3}{\frac{9}{2}+n}
\right)
\right.
\nonumber\\
&\qquad\qquad\qquad\left.
+
c_j^{(0)} 
\left(
\frac{\Delta_0}{(\frac{3}{2}+n)^2}
-
\frac{\Delta_1}{(\frac{5}{2}+n)^2}
+
\frac{\Delta_2}{(\frac{7}{2}+n)^2}
-
\frac{\Delta_3}{(\frac{9}{2}+n)^2}
\right)
\right\}
\nonumber\\
&-2\frac{\alpha_s}{\pi}
C_A(1-\delta)
\sum_{n=0}^\infty 
\sum_{j=0}^n
\delta ^{\frac{1}{2}+n}
\left\{
\tilde c_j
\left[
\psi \left(\frac{3}{2}+n\right)
-
\psi (n-j+1)
\right]
-c_j^{(0)}
\psi '\left(\frac{3}{2}+n\right)
\right\}
\nonumber\\
&
+2\frac{\alpha_s}{\pi}
\left(
C_A\ln \delta +\beta_0
\right)
\sum_{n=0}^\infty 
\delta ^{\frac{1}{2}+n}
{{\tilde c_n}}
+\mathcal{O}(\epsilon)
\end{align}
where 
\begin{eqnarray}
  \Delta_0 &=& -1+3\delta-2\delta^2+\delta^3\,,\nonumber\\
  \Delta_1 &=& \delta^2(2+\delta)\,,\nonumber\\
  \Delta_2 &=& \delta^2(1+2\delta)\,,\nonumber\\
  \Delta_3 &=& \delta^3\,,\nonumber\\
  \tilde c_j&=&
  \frac{c_j^{(0)} }{\epsilon}
  +c_j^{(1)}-c_j^{(0)} \ln \delta 
  +c_j^{(0)} \ln (\mu_r^2/\mu_f^2)\,.
\end{eqnarray}

Our result for the $qg$ channel reads
\begin{align}
\sigma_{qg,\mathrm{coll}}^{(2),n_h^3}
&=
\sigma_{\bar qg,\mathrm{coll}}^{(2),n_h^3}
=
\sigma_{gq,\mathrm{coll}}^{(2),n_h^3}
=
\sigma_{g\bar q,\mathrm{coll}}^{(2),n_h^3}
\nonumber\\
&=\frac{\alpha_s}{\pi}
\frac{C_F}{4}
\sum_{n=0}^\infty 
\sum_{j=0}^n
\delta ^{\frac{3}{2}+n}
\frac{(n-j+2)!}{(n-j)!}
\left\{
c_j^{(0)} 
\left(
\frac{1+\delta^2}{(\frac{3}{2}+n)^2}
-
\frac{2\delta(1+\delta)}{(\frac{5}{2}+n)^2}
+
\frac{2\delta^2}{(\frac{7}{2}+n)^2}
\right)
\right.
\nonumber\\
&\qquad\left.
+
\tilde c_j
\left(
\frac{1+\delta^2}{\frac{3}{2}+n}
-
\frac{2\delta(1+\delta)}{\frac{5}{2}+n}
+
\frac{2\delta^2}{\frac{7}{2}+n}
\right)
\right\}
+\mathcal{O}(\epsilon)
\,.
\end{align}
To obtain terms to $\delta^{1/2+N}$ one should evaluate the series
representations up to $n=N$, and then discard any incomplete higher-order terms
which are produced.

In the ancillary file~\cite{progdata} we present expressions for
$\sigma_{gg}^{(1),n_h^3}$ (from which the coefficients $c_n^{(0)}$
and $c_n^{(1)}$ can be extracted),
$\sigma_{gg,\rm coll}^{(2),n_h^3}$ and $\sigma_{qg,\rm coll}^{(2),n_h^3}$
for arbitrary renormalization and factorization scale and expanded up
to $\delta^{199/2}$ and $\rho^4$. In order to illustrate the
structure of our result we provide some leading non-vanishing terms in
the $\delta$ and $\rho$ expansion which are given by
\begin{eqnarray}
  \sigma_{gg,\rm coll}^{(2),n_h^3} 
  &=&
\frac{\api^4 \GF^2 \Mh^2 \nh^3}{\pi } \Bigg\{
	\cAC \ItRC \sqrt{\delta} \bigg(
		-\frac{\log(\delta)}{432}
		+\frac{13}{5184}
		-\frac{\log{2}}{216}
	\bigg)
	+\frac{\nl \ItRC^2 \sqrt{\delta}}{1296}
\nonumber\\&& \evenspace
	+\rr \Bigg[
		\cAC \ItRC \bigg(
			\frac{\sqrt{\delta}}{\ep} \bigg\{-\frac{7 \log(\delta)}{51840}+\frac{91}{622080}-\frac{7 \log{2}}{25920}\bigg\}
			+\sqrt{\delta} \bigg\{
				\log(\delta) \bigg[
					-\frac{7 \lmuMh}{51840}
\nonumber\\&& \evenspace
					-\frac{7 \lmuMt}{17280}
					-\frac{7 \lmumuf}{51840}
					-\frac{451}{622080}
					+\frac{7 \log{2}}{12960}
				\bigg]
				+\frac{7 \log^2(\delta)}{51840}
				+\frac{91 \lmuMh}{622080}
				-\frac{7 \lmuMh \log{2}}{25920}
\nonumber\\&& \evenspace
				+\frac{91 \lmuMt}{207360}
				-\frac{7 \lmuMt \log{2}}{8640}
				+\frac{91 \lmumuf}{622080}
				-\frac{7 \lmumuf \log{2}}{25920}
				-\frac{7 \pi^2}{103680}
				+\frac{121}{103680}
\nonumber\\&& \evenspace
				+\frac{7 \log^2(2)}{12960}
				-\frac{451 \log{2}}{311040}
			\bigg\}
		\bigg)
		+\nl \ItRC^2 \bigg(
			\frac{7 \sqrt{\delta}}{155520 \ep}
			+\sqrt{\delta} \bigg\{
				-\frac{7 \log(\delta)}{155520}
				+\frac{7 \lmuMh}{155520}
\nonumber\\&& \evenspace
				+\frac{7 \lmuMt}{51840}
				+\frac{7 \lmumuf}{155520}
				+\frac{1}{5184}
				-\frac{7 \log{2}}{77760}
			\bigg\}
		\bigg)
	\Bigg]
	+ \mathcal{O}(\rr^2)
	+ \mathcal{O}(\delta^{3/2})
\Bigg\}\,,
  \label{eq::nh3_coll_gg}\\
  \sigma_{qg,\rm coll}^{(2),n_h^3} 
  &=& \sigma_{\bar{q}g,\rm coll}^{(2),n_h^3} \,\,=\,\,
      \sigma_{gq,\rm coll}^{(2),n_h^3} \,\,=\,\, \sigma_{g\bar{q},\rm coll}^{(2),n_h^3}
      \nonumber\\
      &=&
      \frac{\api^4 \GF^2 \Mh^2 \nh^3}{\pi } \cRC \ItRC \Bigg\{
	-\frac{\delta^{3/2}}{2592}
	+ \rr \Bigg[
		-\frac{7 \delta^{3/2}}{311040 \ep}
		+\delta^{3/2} \bigg(
			\frac{7 \log(\delta)}{311040}
			-\frac{7 \lmuMh}{311040}
\nonumber\\&& \evenspace
			-\frac{7 \lmuMt}{103680}
			-\frac{7 \lmumuf}{311040}
			-\frac{13}{116640}
			+\frac{7 \log{2}}{155520}
		\bigg)
	\Bigg]
	+ \mathcal{O}(\rr^2)
	+ \mathcal{O}(\delta^{5/2})
\Bigg\}
      \,,
\end{eqnarray}
where $\cAC=3$, $\cRC=4/3$ and $\ItRC=1/2$ are colour factors. Furthermore,
we have introduced the notation
\begin{eqnarray}
  a_s=\frac{\alpha_s^{(5)}(\mu_r)}{\pi}\,,\quad
  \lmuMt=\log\frac{\mu_r^2}{M_t^2}\,,\quad
  \lmuMh=\log\frac{\mu_r^2}{m_H^2}\,,\quad
  \lmumuf=\log\frac{\mu_r^2}{\mu_f^2}\,.
\end{eqnarray}


\subsection{\label{sub::workflow}Workflow to compute $\sigma_{ij,\rm real}^{(2),n_h^3}$}

To obtain the NNLO $n_h^3$ real-virtual contributions, we have to
consider five-loop forward-scattering amplitudes with three closed top
quark loops, each of which is coupled to one or two Higgs
bosons. Since an exact calculation is currently not possible we
perform an asymptotic expansion (see, e.g.,
Ref.~\cite{Smirnov:2002pj}) for $M_t^2\gg m_H^2, s$.  As a result, we
obtain products of three one-loop vacuum integrals with a two-loop
integral with two or three massive Higgs boson propagators and forward
scattering kinematics. For the latter we have to compute the imaginary
part involving two Higgs bosons and a gluon or \mbox{(anti-)quark}.

The expansion in $1/M_t$ quickly develops a large number of terms. For
this reason we pre-compute the $1/M_t$ expansion of the one-loop
vacuum integrals with one or two external Higgs bosons and two or
three external gluons.  They are then inserted into the two-loop
diagrams which are generated with effective Higgs-gluon vertices.

In the following we provide more technical details,
the description applies to both NLO and NNLO contributions.
We generate the one- and two-loop diagrams using {\tt
  qgraf}~\cite{Nogueira:1991ex} and select only the diagrams containing the
relevant cuts using additional scripts. This output is then processed
by {\tt q2e} and {\tt
  exp}~\cite{Harlander:1997zb,Seidensticker:1999bb,q2eexp}, which
generate {\tt FORM}~\cite{Ruijl:2017dtg} code for the amplitudes and
map them onto the corresponding integral families. We compute the
colour factors of the diagrams using {\tt
  color}~\cite{vanRitbergen:1998pn}. The tadpole integrals of the
``effective vertices'' are evaluated using {\tt
  MATAD}~\cite{Steinhauser:2000ry}.

We initially define the one- and two-loop integral families without
specifying forward-scattering kinematics, but rather with three independent
(incoming) external momenta, $q_1, q_2, q_3$ and the relation
$q_4 = -q_1 -q_2 -q_3$. The identification $q_2=-q_3$ and $q_1=-q_4$ is
applied at a later stage (see below).

We use this setup to obtain scalar expressions for each amplitude after
averaging over the polarizations, spins and colours.  Afterwards we
decompose scalar products in the numerators in terms of denominator factors
and map the scalar integrals onto 1~one-loop and 50~two-loop
four-point integral families which are characterized by four and nine
indices, respectively. In the two-loop case seven indices correspond
to propagators and two to irreducible numerators, which we write
as inverse propagators.

\begin{figure}[t]
  \begin{center}
    \includegraphics[width=.5\textwidth]{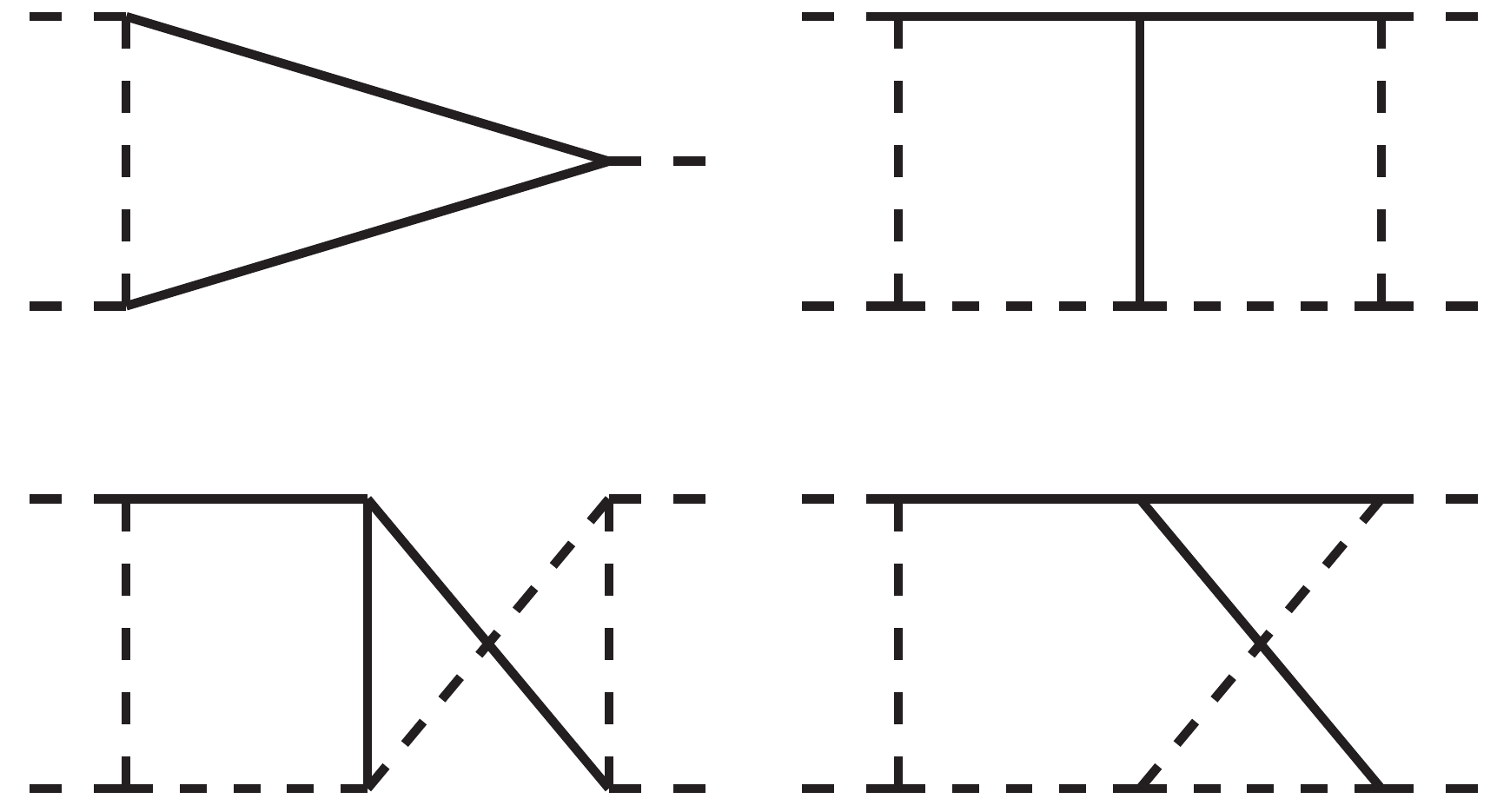}
  \end{center}
  \caption{\label{fig::fam_fire} One- and two-loop integral families
    with forward scattering kinematics. Solid and dashed lines
    represent massive and massless propagators. There are three more
    families at two loops.
    Their master integrals can be mapped to the three families
    shown in the figure.}
\end{figure}

\begin{figure}[t]
  \begin{center}
    \includegraphics[width=.5\textwidth]{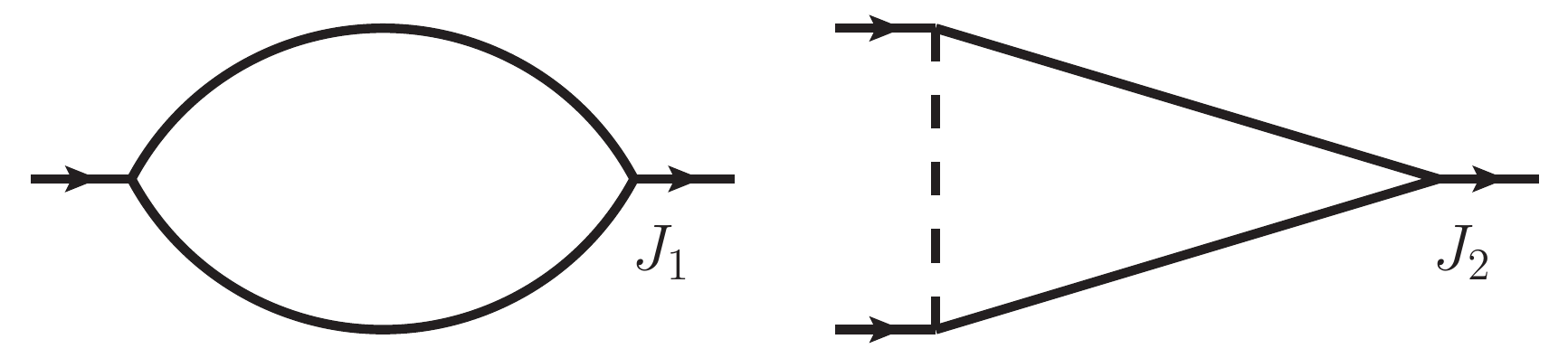}
  \end{center}
  \caption{\label{fig::masters1l}The one-loop master integrals. 
    Solid and dashed lines represent massive and
    massless propagators. It is understood that the momenta $q_1$ and
    $q_2$ enter the diagrams on the left in the upper and lower lines,
    respectively.} 
\end{figure}

\begin{figure}[t]
  \begin{center}
    \includegraphics[width=\textwidth]{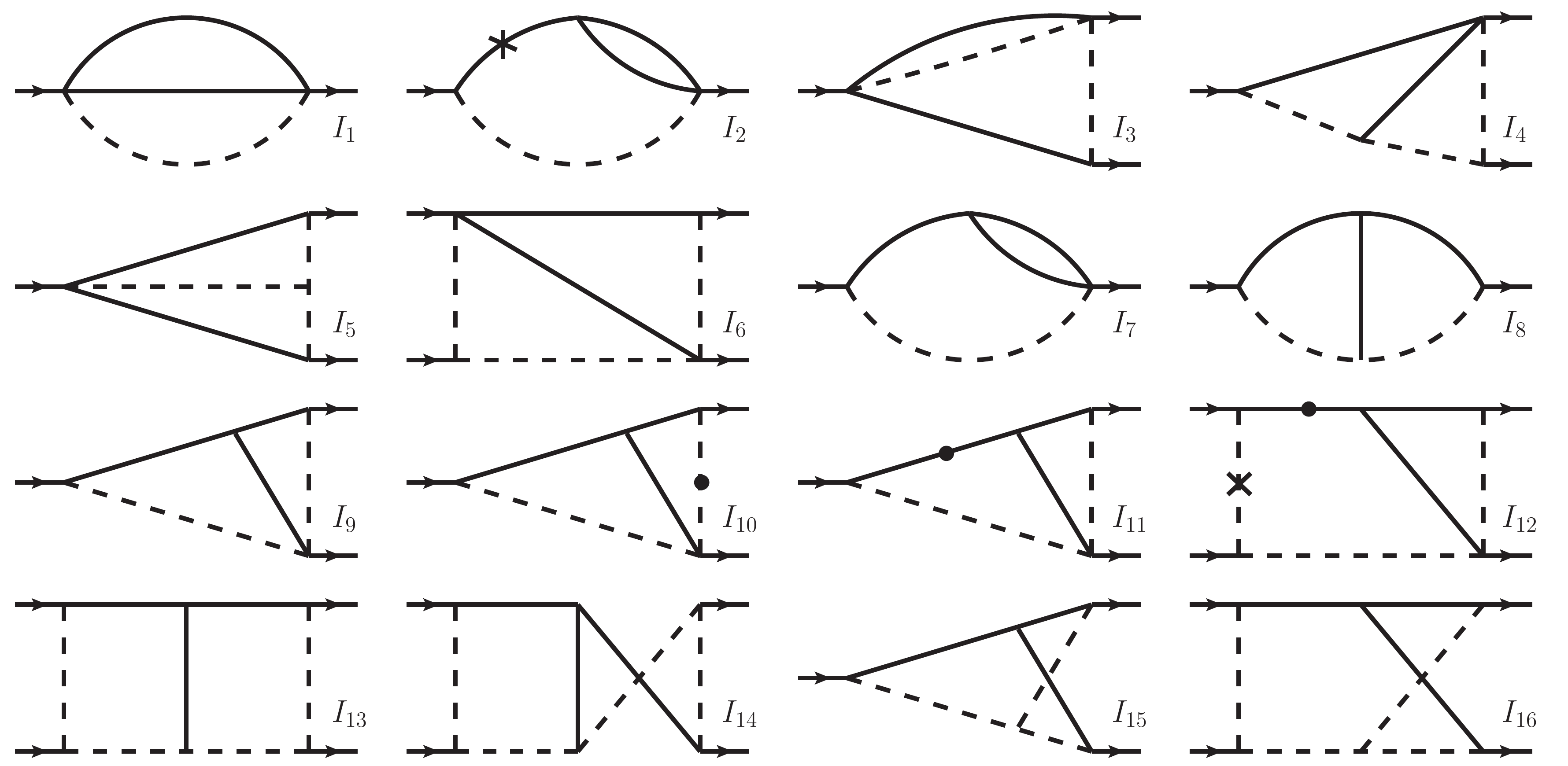}
  \end{center}
  \caption{\label{fig::masters}
    Minimal set of two-loop
    master integrals.  Solid and dashed lines represent massive and
    massless propagators. It is understood that the momenta $q_1$ and
    $q_2$ enter the diagrams on the left and leave them on the right
    in the upper and lower lines, respectively.}
\end{figure}

Next we impose the forward scattering kinematics, i.e. we set
$q_3 = -q_2$ and $q_4 = -q_1$. This results in the propagators becoming
linearly dependent.  After partial fraction decomposition and identification
of identical families we are left with 1~one-loop family of four propagators and 6~two-loop
families of seven propagators.  Graphical representations are given
in Fig.~\ref{fig::fam_fire}.
These integral families are suitable for
IBP reduction which we perform with the help of {\tt
  FIRE5}~\cite{Smirnov:2014hma}.  We use {\tt FindRules}, a built-in
command of {\tt FIRE}, to identify master integrals of different
families and arrive at 2~one-loop and 16~two-loop master integrals
which are depicted in Figs.~\ref{fig::masters1l}
and~\ref{fig::masters}.

The master integrals depend on $x=m_H^2/s$ where $0 < x < 1/4$. To
obtain analytic results we use the powerful method of differential
equations~\cite{Kotikov:1990kg,Kotikov:1991hm,Kotikov:1991pm}, which
we derive with the help of {\tt LiteRed}~\cite{Lee:2012cn,Lee:2013mka}.
To obtain the solutions we proceed in two ways.  In the first approach we
adopt the idea of Ref.~\cite{Grigo:2013rya} and expand around the
threshold (i.e. $x=1/4$).
We compute boundary conditions for all master integrals for
$\delta\to0$ and then use the differential equations to obtain for each master
integral an expansion up to $\delta^{1019/2}$.  More detail on this method
can be found in Appendix~\ref{app::MI_delta}.

In the second approach we compute the master integrals without expanding in
$\delta$ and express the result in terms of Goncharov
polylogarithms~\cite{Goncharov:1998kja}.  The boundary conditions can be taken
over from the first approach. We provide more details in
Appendix~\ref{app::MI_exact}.  As we show in Appendix~\ref{app::MI_exact} it
is sufficient to use the (mathematically simpler) $\delta$-expanded results
for phenomenological applications.

In the following we illustrate the structure of our results and
provide the leading expansion terms in $\rho$ and $\delta$ for $\sigma_{ij,\rm
  real}^{(2),n_h^3}$:
\begin{eqnarray}
  \sigma_{gg,\rm real}^{(2),n_h^3} 
  &=&
\frac{\api^4 \GF^2 \Mh^2 \nh^3}{\pi } \cAC \ItRC \Bigg\{
	-\frac{\sqrt{\delta}}{864 \ep}
	+\sqrt{\delta} \bigg(\frac{\log(\delta)}{288}-\frac{\lmuMh}{432}-\frac{\lmuMt}{288}-\frac{7}{864}+\frac{\log{2}}{108}\bigg)
\nonumber\\&& \evenspace
	+ \rr \Bigg[
		-\frac{7 \sqrt{\delta}}{103680 \ep^2}
		+\frac{\sqrt{\delta}}{\ep} \bigg(
			\frac{7 \log(\delta)}{34560}
			-\frac{7 \lmuMh}{51840}
			-\frac{7 \lmuMt}{34560}
			-\frac{29}{51840}
			+\frac{7 \log{2}}{12960}
		\bigg)
\nonumber\\&& \evenspace
		+\sqrt{\delta} \bigg(
			\log(\delta) \bigg\{
				\frac{7 \lmuMh}{17280}
				+\frac{7 \lmuMt}{11520}
				+\frac{29}{17280}
				-\frac{7 \log{2}}{4320}
			\bigg\}
			-\frac{7 \log^2(\delta)}{23040}
			-\frac{7 \lmuMh^2}{51840}
\nonumber\\&& \evenspace
			-\frac{7 \lmuMh \lmuMt}{17280}
			-\frac{29 \lmuMh}{25920}
			+\frac{7 \lmuMh \log{2}}{6480}
			-\frac{7 \lmuMt^2}{23040}
			-\frac{29 \lmuMt}{17280}
			+\frac{7 \lmuMt \log{2}}{4320}
\nonumber\\&& \evenspace
			+\frac{161 \pi^2}{1244160}
			-\frac{61}{17280}
			-\frac{7 \log^2(2)}{3240}
			+\frac{29 \log{2}}{6480}
		\bigg)
	\Bigg]
	+ \mathcal{O}(\rr^2)
	+ \mathcal{O}(\delta^{3/2})
\Bigg\}\,,
      \label{eq::nh3_real_gg}\\
  \sigma_{qg,\rm real}^{(2),n_h^3} 
  &=& \sigma_{\bar{q}g,\rm real}^{(2),n_h^3} \,\,=\,\,
      \sigma_{gq,\rm real}^{(2),n_h^3} \,\,=\,\, \sigma_{g\bar{q},\rm real}^{(2),n_h^3}
      \nonumber\\
  &=&
\frac{\api^4 \GF^2 \Mh^2 \nh^3}{\pi } \cRC \ItRC \Bigg\{
	\frac{\delta^{3/2}}{2592}
	+ \rr \Bigg[
		\frac{7 \delta^{3/2}}{311040 \ep}
		+\delta^{3/2} \bigg(
			-\frac{7 \log(\delta)}{103680}
			+\frac{7 \lmuMh}{155520}
\nonumber\\&& \evenspace
			+\frac{7 \lmuMt}{103680}
			+\frac{13}{62208}
			-\frac{7 \log{2}}{38880}
		\bigg)
	\Bigg]
	+ \mathcal{O}(\rr^2)
	+ \mathcal{O}(\delta^{5/2})
\Bigg\}\,,
      \\
  \sigma_{q\bar{q},\rm real}^{(2),n_h^3} 
  &=&
\frac{\api^4 \GF^2 \Mh^2 \nh^3}{\pi } \cRC^2 \Bigg\{
	-\frac{2 \rr \delta^{9/2}}{76545}
	-\frac{128 \rr^2 \delta^{9/2}}{2679075}
	+ \mathcal{O}(\rr^3)
	+ \mathcal{O}(\delta^{11/2})
\Bigg\}
\,.
\end{eqnarray}

We have managed to perform the expansion up to order $\rho^4$ in the
Feynman gauge.
We additionally perform the expansion up to order $\rho^2$ for a general
QCD gauge parameter $\xi$, and find that all dependence on $\xi$ drops
out after summing the contributions of all bare diagrams. This provides
a strong check of our calculation.



\section{Results\label{sec:res}}

We start by presenting analytic results for the partonic cross
sections $\sigma_{gg}^{(2),n_h^3}$. Combining
Eqs.~(\ref{eq::nh3_coll_gg}) and~(\ref{eq::nh3_real_gg}) with the
virtual corrections from~\cite{Grigo:2015dia,DavSte19} the first two
expansion terms in $\rho$ and $\delta$ are given by
\begin{eqnarray}
  \sigma_{gg}^{(2),n_h^3} &=&
\frac{\api^4 \GF^2 \Mh^2 \nh^3}{\pi } \Bigg\{
	\delta^{3/2} \bigg(
		\log(\delta) \bigg\{
			\frac{\lmuMh}{432}
			+\frac{1}{81}
			-\frac{\log{2}}{72}
		\bigg\}
		-\frac{1}{432} \log^2(\delta)
		-\frac{25 \lmuMh}{3456}
\nonumber\\&& \evenspace
		+\frac{1}{216} \lmuMh \log{2}
		+\frac{5 \pi^2}{82944}
		-\frac{2053}{62208}
		-\frac{\log^2(2)}{54}
		+\frac{\log{2}}{27}
	\bigg)
	+\nl \delta^{3/2} \bigg(
		\frac{\lmuMh}{15552}
\nonumber\\&& \evenspace
		+\frac{5}{46656}
	\bigg)
	+ \rr \Bigg[
		\sqrt{\delta} \bigg(
			\log(\delta) \bigg\{\frac{7 \lmuMh}{34560}+\frac{7}{8640}-\frac{7 \log{2}}{5760}\bigg\}
			-\frac{7 \log^2(\delta)}{34560}
			-\frac{413 \lmuMh}{829440}
\nonumber\\&& \evenspace
			+\frac{7 \lmuMh \log{2}}{17280}
			+\frac{7 \pi^2}{1327104}
			-\frac{30587}{14929920}
			-\frac{7 \log^2(2)}{4320}
			+\frac{7 \log{2}}{2880}
		\bigg)
\nonumber\\&& \evenspace
		+ \delta^{3/2} \bigg(
			\log(\delta) \bigg\{\frac{19 \lmuMh}{103680}+\frac{283}{311040}-\frac{19 \log{2}}{17280}\bigg\}
			-\frac{19 \log^2(\delta)}{103680}
			+\frac{121 \lmuMt}{622080}
\nonumber\\&& \evenspace
			-\frac{2077 \lmuMh}{2488320}
			+\frac{19 \lmuMh \log{2}}{51840}
			+\frac{10549 \zeta_3}{31850496}
			-\frac{1109 \pi^2}{59719680}
			-\frac{32155177}{10749542400}
\nonumber\\&& \evenspace
			-\frac{19 \log^2(2)}{12960}
			+\frac{19 \log{2}}{6480}
		\bigg)
		+\nl \bigg(
			\sqrt{\delta} \bigg\{
				\frac{7 \lmuMh}{1244160}
				+\frac{7}{746496}
			\bigg\}
			+ \delta^{3/2} \bigg\{
				\frac{19 \lmuMh}{3732480}
\nonumber\\&& \evenspace
				+\frac{11}{11197440}
			\bigg\}
		\bigg)
	\Bigg]
	+ \mathcal{O}(\rr^2)
	+ \mathcal{O}(\delta^{5/2})
\Bigg\}\,.
\end{eqnarray}

The results for $\sigma_{ij}^{(2),n_h^3}$ for $\rho=0$ have been obtained
for the first time in Ref.~\cite{deFlorian:2013jea} (denoted by
$\hat{\sigma}^b$ in that paper). Note, however, that the final phase-space
integration has been performed numerically and results are presented for hadronic
quantities, and therefore a direct comparison with our results is non-trivial.

\begin{figure}
    \begin{center}
      \begin{tabular}{cc}
        \includegraphics[width=0.52\textwidth]{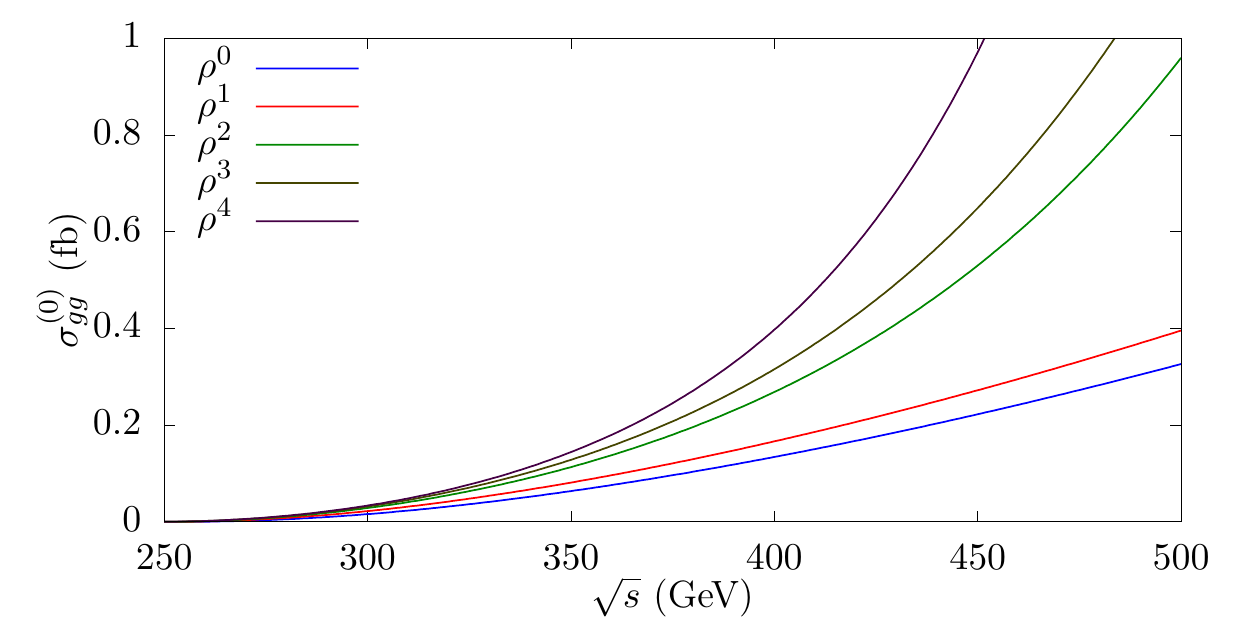}  &
        \includegraphics[width=0.52\textwidth]{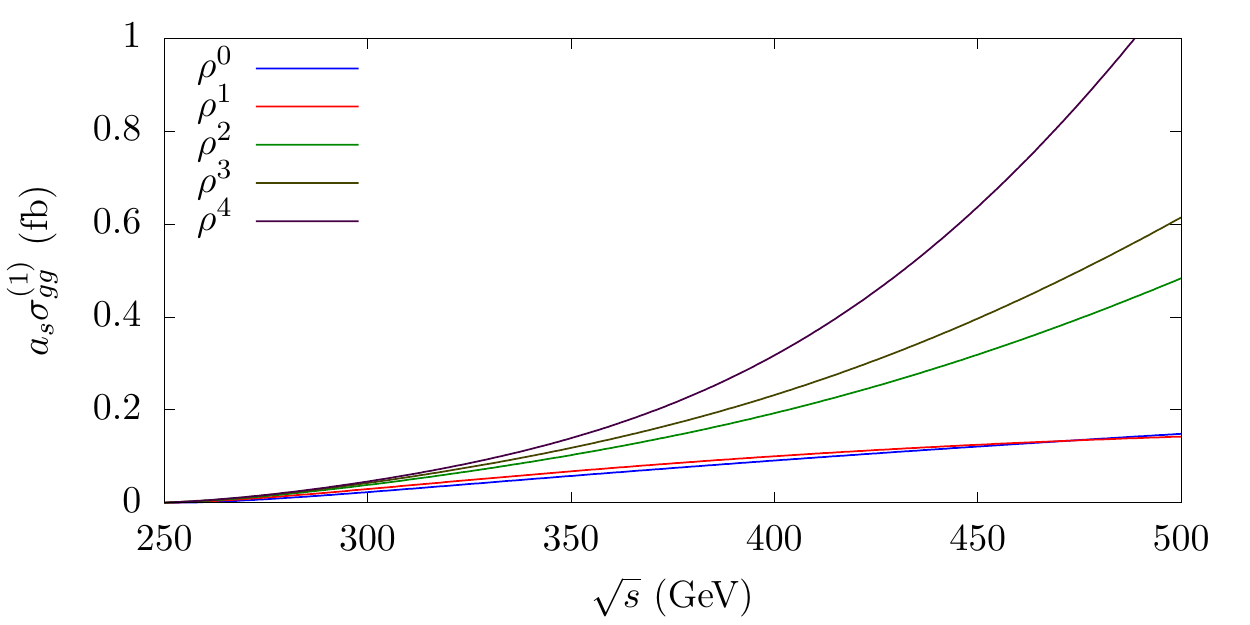}
        \\
        \includegraphics[width=0.52\textwidth]{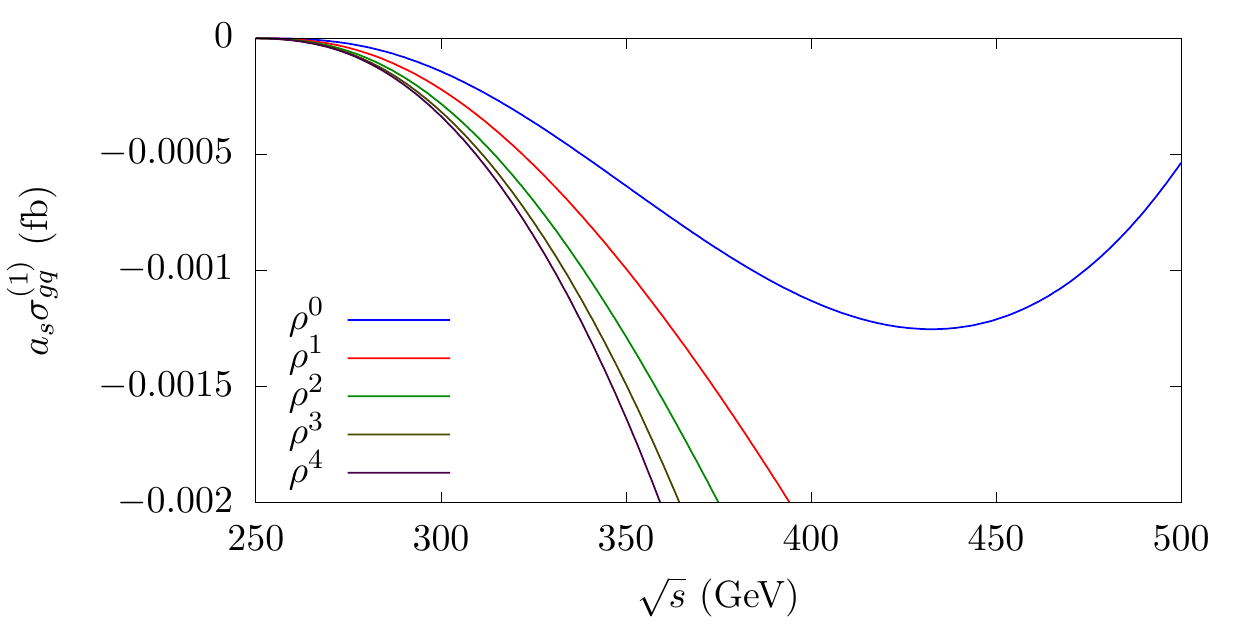} &
        \includegraphics[width=0.52\textwidth]{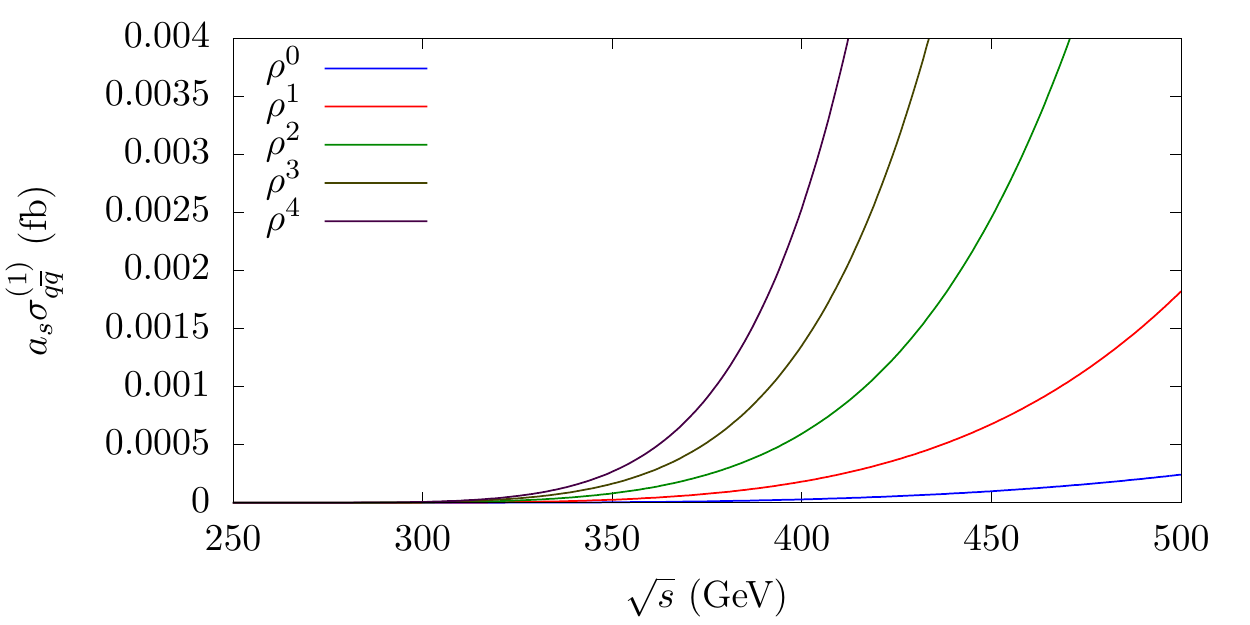}
        \\
        \includegraphics[width=0.52\textwidth]{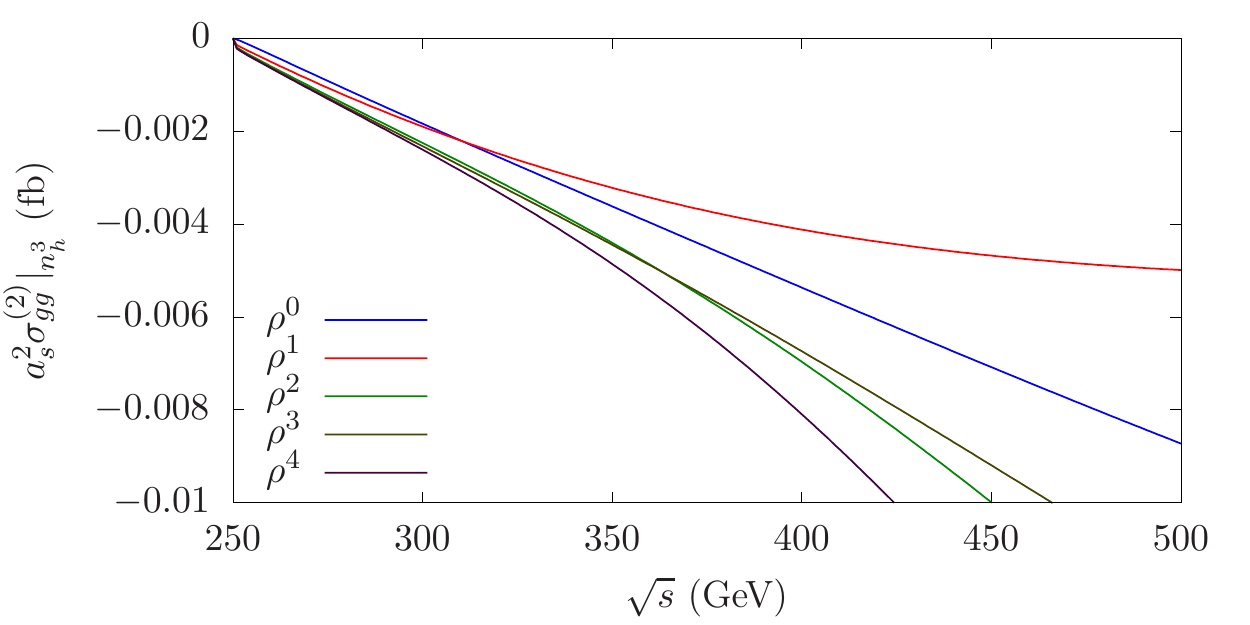}&
        \includegraphics[width=0.52\textwidth]{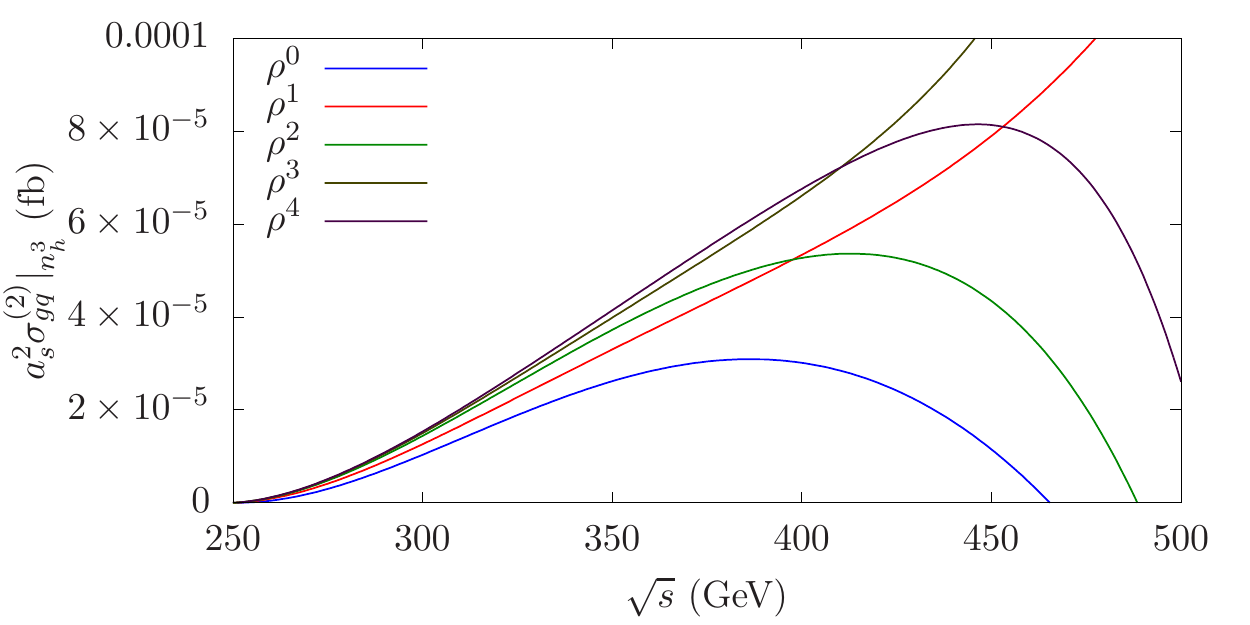}
        \\
        \includegraphics[width=0.52\textwidth]{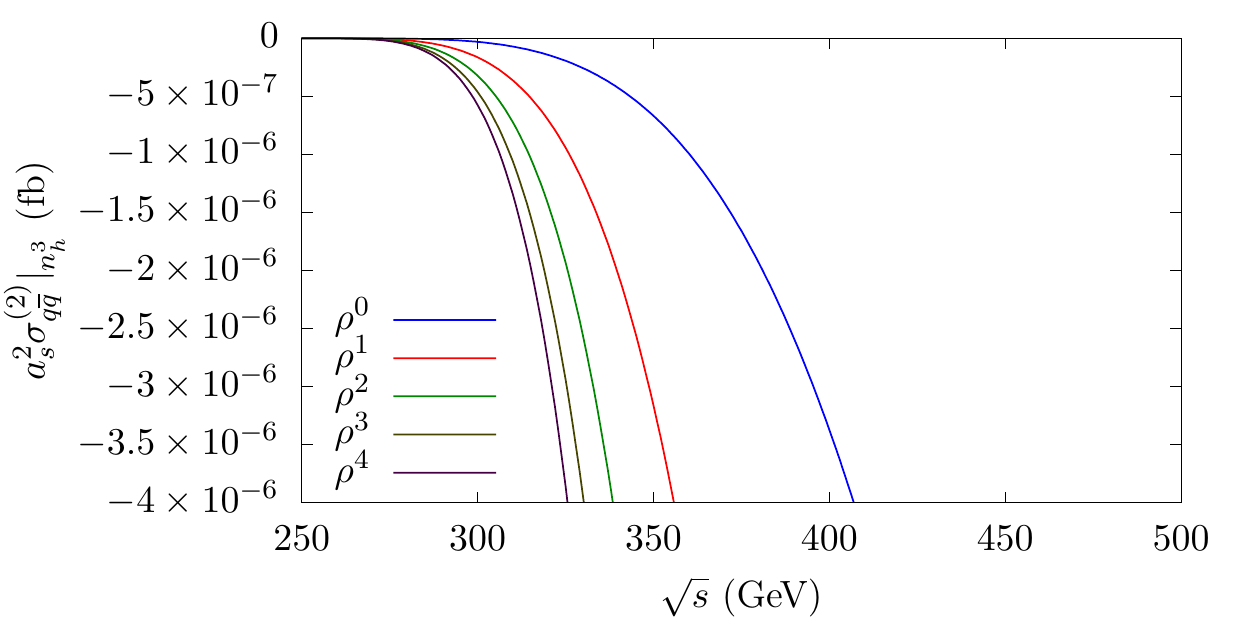}
      \end{tabular}
      \caption{\label{fig::sig_part}LO, NLO and NNLO partonic cross sections
        as a function of $\sqrt{s}$ with $a_s=\alpha_s^{(5)}(m_H)/\pi$.}
    \end{center}
\end{figure}

In Fig.~\ref{fig::sig_part} we show the partonic cross sections as a function
of $\sqrt{s}$. In such situations, the exact LO contribution is often factored
out in order to improve the behaviour at high energies. In
Fig.~\ref{fig::sig_part} we refrain from doing so since we want to
illustrate the convergence properties below the top quark pair threshold.  For
convenience we repeat the well-known LO and NLO results~\cite{Grigo:2013rya}.

The $gg$ initiated NNLO contribution shows a similar pattern as at LO and NLO.
Up to $\sqrt{s}\approx 330$~GeV a reasonable convergence is observed when
including higher order terms in $\rho$. Beyond the top threshold we have
no convergence.
The $qg$ channel also shows good convergence up to the top quark threshold 
both at NLO and NNLO. No sign of convergence is observed for the $q\bar{q}$
channel. Note, however, that the contributions from  production channels
with quarks in the initial state are significantly smaller than the $gg$ channel.

At NLO the $n_h^3$ contribution is numerically much smaller than the
remaining parts~\cite{Degrassi:2016vss}.
We thus expect that also at NNLO the $n_h^2$ terms will be numerically more
important.

Let us mention that based on previous
experience~\cite{Grigo:2013rya,Grigo:2015dia}, we did not expect
better convergence behaviour than what is shown in
Fig.~\ref{fig::sig_part}.  Nonetheless, the higher order terms
in $\rho$ are important ingredients for the construction of
approximations. For example, at NLO, the $\rho^3$ and $\rho^4$
terms help to obtain stable results
for the cross section when combining
large-$M_t$ results with information about the threshold behaviour, using
Pad\'e approximants~\cite{Grober:2017uho}.


\section{Conclusions\label{sec:conclusion}}

We compute real radiation corrections to the cross section $gg\to HH$ by
applying the optical theorem to Feynman diagrams with forward scattering
kinematics. We concentrate on the subset of Feynman diagrams which involve
three closed top quark loops, each of which couples to either one or two
Higgs bosons, the so-called $n_h^3$ contribution.  With the help of an
asymptotic expansion for large top quark masses the five-loop diagrams
factorize into three one-loop vacuum integrals and two-loop phase-space
integrals which depend on $s$ and $m_H^2$.  After IBP reducing the integral families
we are left with 16 master integrals, which we compute both as an expansion
around the threshold and as an exact expression in $s/m_H^2$.  Although the radius of convergence
is limited to a small region around threshold (i.e. $s\approx 4m_H^2$) our
results are useful to provide information about the reliability of the
effective-theory result. Furthermore, we expect that the power-suppressed top
mass corrections will prove to be useful ingredients for approximation procedures and,
last but not least, they are important benchmarks for future numerical
calculations.

Work to compute the remaining ($n_h^2$) contributions to the NNLO real corrections
to double Higgs boson production is ongoing, and will be published in a future paper.


\section*{Acknowledgements}

F.H. thanks Max Delto for fruitful discussions and acknowledges the
support of the DFG-funded Doctoral School KSETA.
This research was supported by the Deutsche Forschungsgemeinschaft (DFG,
German Research Foundation) under grant 396021762 --- TRR 257
``Particle Physics Phenomenology after the Higgs Discovery''.

\appendix


\section{\label{app::MI_delta}Threshold expansion of master integrals}

In this appendix we describe the calculation of the phase-space master
integrals as an expansion around the threshold, i.e., for small values of
$\delta$, see Eq.~(\ref{eq::delta}).
For simplicity, we denote the Higgs boson mass by $m$ (instead of $m_H$).

All 16 master integrals (cf. Fig.~\ref{fig::masters}) can be expressed in the form
\begin{align}
  I_i=\left(\frac{e^{\gamma_E}\mu_r^2}{4\pi}\right)^{2\epsilon} \int 
  \mathcal{D}p_3
  \mathcal{D}p_4
  \mathcal{D}p_5
  (2\pi)^d\delta ^{(d)} (q_1+q_2-p_3-p_4-p_5)
  \mathcal{Q}_i
  \,,
  \label{e1}
\end{align}
where the momenta $p_3$ and $p_4$ correspond to massive (Higgs) particles 
and the integration measures are given by
\begin{align}
\mathcal{D}p_j
\equiv
\frac{\mathrm{d}^{d-1}p_j}{(2\pi)^{d-1}}
\frac{1}{2E_j}
=
\frac{p_j^{d-2}\mathrm{d}p_j}{(2\pi)^{d-1}}
\frac{1}{2E_j}
\mathrm{d}\Omega^{(j)}_{d-1}
\qquad \mathrm{for}\quad j=3,4
\end{align}
with $E_j\equiv \sqrt{m^2+|\vec p_j|^2}$.
The momentum $p_5$ corresponds to massless final-state parton
and the integration measure reads
\begin{align}
\mathcal{D}p_5
\equiv
\frac{1}{2}
\frac{p_5^{d-3}\mathrm{d}p_5}{(2\pi)^{d-1}}
\mathrm{d}\Omega^{(5)}_{d-1}
\,.
\end{align}

The quantities $Q_i$ in Eq.~(\ref{e1}) are given by
\begin{align}
&
~\mathcal{Q}_{1}=1,~\mathcal{Q}_{2}=m^2-(p_3+p_4)^2,~\mathcal{Q}_{3}=\frac{-1}{(q_2-p_4)^2},
\nonumber\\ &
~\mathcal{Q}_{4}=\frac{1}{(q_1-p_5)^2 (p_3+p_5)^2},~\mathcal{Q}_{5}=\frac{1}{(q_1-p_3)^2 (q_2-p_4)^2},
~\mathcal{Q}_{6}=\frac{1}{(q_1-p_5)^2 (q_2-p_4)^2},
\nonumber\\ &
~\mathcal{Q}_{7}=\frac{1}{m^2-(p_3+p_4)^2},
~\mathcal{Q}_{8}=\frac{-1}{(p_3+p_5)^2 \left(m^2-(p_3+p_4)^2\right)},
\nonumber\\ &
~\mathcal{Q}_{9}=\frac{-1}{(q_2-p_4)^2 \left(m^2-(p_3+p_4)^2\right)},
~\mathcal{Q}_{10}=\frac{1}{(q_2-p_4)^4 \left(m^2-(p_3+p_4)^2\right)},
\nonumber\\ &
~\mathcal{Q}_{11}=\frac{-1}{(q_2-p_4)^2 \left(m^2-(p_3+p_4)^2\right)^2},
~\mathcal{Q}_{12}=\frac{(q_1-p_5)^2}{(q_2-p_4)^2 \left(m^2-(p_3+p_4)^2\right)^2},
\nonumber\\ &
~\mathcal{Q}_{13}=\frac{-1}{(q_1-p_5)^2 (q_2-p_4)^2 (p_3+p_5)^2 \left(m^2-(p_3+p_4)^2\right)},
\nonumber\\ &
~\mathcal{Q}_{14}=\frac{-1}{(q_2-p_4)^2 (q_2-p_5)^2 (p_3+p_5)^2 \left(m^2-(p_3+p_4)^2\right)},
\nonumber\\ &
~\mathcal{Q}_{15}=\frac{1}{(q_1-p_3)^2 (q_2-p_4)^2 \left(m^2-(p_3+p_4)^2\right)},
\nonumber\\ &
~\mathcal{Q}_{16}=\frac{-1}{(q_1-p_3)^2 (q_2-p_4)^2 (q_2-p_5)^2 \left(m^2-(p_3+p_4)^2\right)}
\,.
\end{align}

We parametrise the $d$-dimensional momenta in Eq.~\eqref{e1} as
\begin{align}
&
q_1=\frac{\sqrt{s}}{2}
\left(
\begin{array}{c}
1\\ 0\\ \vdots\\ 0\\ 1
\end{array}
\right)
\!,\:
q_2=\frac{\sqrt{s}}{2}
\left(
\begin{array}{c}
1\\ 0\\ \vdots\\ 0\\ -1
\end{array}
\right)
\!,\:
p_3=
\left(
\begin{array}{c}
E_3\\
\vdots\\
\vdots\\
k\sin\theta_3\cos\phi_3\\
k\cos\theta_3
\end{array}
\right)
\!,\:
p_5=
\left(
\begin{array}{c}
\ell\\
0\\
\vdots\\
0\\
\ell \sin\theta_5\\
\ell \cos\theta_5
\end{array}
\right)
\!,
\end{align}
and use the $\delta$ function in Eq.~\eqref{e1} to express the spacial
components of $p_4$ as $\vec p_4=-\vec p_3-\vec p_5$.
For the $d$-dimensional angular integrations we use
(see, e.g., Ref.~\cite{Somogyi:2011ir})
\begin{align}
\int 
\mathrm{d}\Omega^{(j)}_{d-1}
&=
\frac{2\pi^\frac{d-3}{2}}{\Gamma (\frac{d-3}{2})}
\int ^1_{-1} 
(1-\cos ^2\theta_j)^\frac{d-4}{2}
\mathrm{d}\cos \theta_j 
\int ^1_{-1} 
(1-\cos ^2\phi_j)^\frac{d-5}{2}
\mathrm{d}\cos \phi_j 
\,.
\end{align}

We now consider Eq.~\eqref{e1}, exploit the delta function and arrive at
\begin{align}
I_i
&=
\left(\frac{e^{\gamma_E}\mu_r^2}{4\pi}\right)^{2\epsilon}
\int 
\mathcal{D}p_3
\mathcal{D}p_5
\frac{1}{2E_4}
(2\pi) \delta (\sqrt{s}-E_3-E_4-|\vec p_5|)
\mathcal{Q}_i
\nonumber\\
&=\frac{e^{2\epsilon \gamma_E}}{256\pi^{5-2\epsilon}}
\int \frac{
k^{d-2}\mathrm{d}k
~\mathrm{d}\Omega^{(3)}_{d-1}
~\ell ^{d-3}\mathrm{d}\ell
~\mathrm{d}\Omega^{(5)}_{d-1}
}{E_3E_4}
\nonumber\\
&\quad \times
~\delta (\sqrt{s}-\sqrt{m^2+k^2}-\sqrt{m^2+|\vec k+\vec \ell |^2}-\ell)
\mathcal{Q}_i
\,.
\end{align}
We can now perform the $\ell$-integration by noting that
\begin{align}
&\sqrt{s}-\sqrt{m^2+k^2}-\sqrt{m^2+|\vec k+\vec \ell |^2}-\ell =0\nonumber\\
&\Leftrightarrow
\ell = \ell_\delta
\equiv 
\frac{
\sqrt{s}\left[
-2k^2-2m^2+ k \cos\gamma \left( \sqrt{s} -2\sqrt{k^2+m^2}\right) -\sqrt{s}\sqrt{k^2+m^2}+s\right]}
{2\left[
s+2\sqrt{s}k \cos\gamma -m^2+k^2\left( -1+\cos\gamma ^2\right) 
\right]}
\,,
\label{eq::ldelta}
\end{align}
where $\cos \gamma =\cos \theta_3\cos\theta_5+\sin\theta_3\sin\theta_5\cos \phi_5$.
Thus, we obtain
\begin{align}
I_i
&=\frac{\left(e^{\gamma_E}\mu_r^2\right)^{2\epsilon}}{256\pi^{5-2\epsilon}}
\int_0^{\sqrt{s\delta}/2}
\mathrm{d}k
\int \frac{
\left( 1+\frac{\ell +k\cos \gamma}{\sqrt{m^2+k^2+2k\ell \cos \gamma +\ell^2}} \right)^{-1}
k^{d-2}
~\mathrm{d}\Omega^{(3)}_{d-1}
~\ell ^{d-3}
\mathrm{d}\Omega^{(5)}_{d-1}
~\mathcal{Q}_i
}{\sqrt{m^2+k^2}\sqrt{m^2+k^2+2k\ell \cos \gamma +\ell^2}}
\Big|_{\ell = \ell_\delta}
\,,
\label{eq:app:int}
\end{align}
where the upper limit of the $k$-integration has been determined by the
$\delta$ function, and the factor
$( 1+ [\ell +k\cos \gamma]/\sqrt{m^2+k^2+2k\ell \cos \gamma +\ell^2} )^{-1}$
is the Jacobian of the $\delta$-function.
Up to this point, the expression is exact.
For convenience we
also provide the propagators appearing in the $\mathcal{Q}_i$ in terms of
$m,k,\ell,\delta$,
\begin{align}
(q_2-p_4)^2
=\,&m^2-2m(\sqrt{m^2+k^2+2k\ell \cos \gamma +\ell^2}-k\cos\theta_3-\ell\cos\theta_5)/\sqrt{1-\delta}
\,,\nonumber\\
(q_1-p_3)^2
=\,&m^2-2m(\sqrt{m^2+k^2}+k\cos\theta_3)/\sqrt{1-\delta}
\,,\nonumber\\
(p_3+p_5)^2
=\,&m^2+2(\ell \sqrt{m^2+k^2}-k\ell \cos \gamma)
\,,\nonumber\\
(q_1-p_5)^2
=&-\sqrt{s}\ell (1-\cos\theta_5)
\,,\nonumber\\
(q_2-p_5)^2
=&
-\sqrt{s}\ell (1+\cos\theta_5)
\,,\nonumber\\
m^2-(p_3+p_4)^2
=&-m^2-2(\sqrt{m^2+k^2}\sqrt{m^2+k^2+2k\ell \cos \gamma +\ell^2}+k^2+k\ell\cos\gamma)
\label{a2}
\,.
\end{align}

At this point the expansion in $\delta$ is straightforward.  Since
$k\leq \sqrt{s\delta}/2 = m\sqrt{\delta/(1-\delta)}$ and
$\ell_\delta \le m\delta$, which follow from Eqs.~(\ref{eq:app:int})
and~(\ref{eq::ldelta}) respectively, we can expand in both variables.  In
practice we proceed as follows: after substituting Eq.~(\ref{eq::ldelta}) into
Eq.~(\ref{eq:app:int}) and making a change of integration variable
$k=\xi\sqrt{s\delta}/2$ the resulting integrand is a polynomial in all
integration variables $\xi, \cos\theta_3, \cos\theta_5$ and $\cos\phi$.

We now show, for each master integral, three $\delta$-expansion terms for the
leading coefficient in the $\epsilon$ expansion. Our results read
\begin{align}
I_{1}&=
\mathcal{N}^2 s\left(
\frac{\delta ^{5/2}}{480 \pi ^3}
+\frac{\delta ^{7/2}}{1680 \pi ^3}
+\frac{\delta ^{9/2}}{3360 \pi ^3}
\right)+\mathcal{O}(\delta^{11/2})
+\mathcal{O}(\epsilon^{1})
,\nonumber\\
I_{2}&=
\mathcal{N}^2 s^2\left(
-\frac{\delta ^{5/2}}{640 \pi ^3}
+\frac{\delta ^{7/2}}{4480 \pi ^3}
+\frac{\delta ^{9/2}}{40320 \pi ^3}
\right)+\mathcal{O}(\delta^{11/2})
+\mathcal{O}(\epsilon^{1})
,\nonumber\\
I_{3}&=
\mathcal{N}^2\left(
\frac{\delta ^{5/2}}{120 \pi ^3}
+\frac{\delta ^{7/2}}{280 \pi ^3}
+\frac{11 \delta ^{9/2}}{3780 \pi ^3}
\right)+\mathcal{O}(\delta^{11/2})
+\mathcal{O}(\epsilon^{1})
,\nonumber\\
I_{4}&=
\frac{\mathcal{N}^2}{\epsilon s}\left(
\frac{\delta ^{3/2}}{48 \pi ^3}
+\frac{\delta ^{5/2}}{120 \pi ^3}
+\frac{13 \delta ^{7/2}}{1680 \pi ^3}
\right)+\mathcal{O}(\delta^{9/2})
+\mathcal{O}(\epsilon^{0})
,\nonumber\\
I_{5}&=
\frac{\mathcal{N}^2}{s}\left(
\frac{\delta ^{5/2}}{30 \pi ^3}
+\frac{4 \delta ^{7/2}}{105 \pi ^3}
+\frac{37 \delta ^{9/2}}{945 \pi ^3}
\right)+\mathcal{O}(\delta^{11/2})
+\mathcal{O}(\epsilon^{1})
,\nonumber\\
I_{6}&=
\frac{\mathcal{N}^2}{\epsilon s}\left(
-\frac{\delta ^{3/2}}{48 \pi ^3}
-\frac{\delta ^{5/2}}{60 \pi ^3}
-\frac{23 \delta ^{7/2}}{1680 \pi ^3}
\right)+\mathcal{O}(\delta^{9/2})
+\mathcal{O}(\epsilon^{0})
,\nonumber\\
I_{7}&=
\mathcal{N}^2\left(
-\frac{\delta ^{5/2}}{360 \pi ^3}
-\frac{\delta ^{7/2}}{504 \pi ^3}
-\frac{11 \delta ^{9/2}}{6804 \pi ^3}
\right)+\mathcal{O}(\delta^{11/2})
+\mathcal{O}(\epsilon^{1})
,\nonumber\\
I_{8}&=
\frac{\mathcal{N}^2}{s}\left(
\frac{\delta ^{5/2}}{90 \pi ^3}
+\frac{2 \delta ^{7/2}}{315 \pi ^3}
+\frac{7 \delta ^{9/2}}{1215 \pi ^3}
\right)+\mathcal{O}(\delta^{11/2})
+\mathcal{O}(\epsilon^{1})
,\nonumber\\
I_{9}&=
\frac{\mathcal{N}^2}{s}\left(
-\frac{\delta ^{5/2}}{90 \pi ^3}
-\frac{\delta ^{7/2}}{105 \pi ^3}
-\frac{26 \delta ^{9/2}}{2835 \pi ^3}
\right)+\mathcal{O}(\delta^{11/2})
+\mathcal{O}(\epsilon^{1})
,\nonumber\\
I_{10}&=
\frac{\mathcal{N}^2}{s^2}\left(
-\frac{2 \delta ^{5/2}}{45 \pi ^3}
-\frac{22 \delta ^{7/2}}{315 \pi ^3}
-\frac{158 \delta ^{9/2}}{1701 \pi ^3}
\right)+\mathcal{O}(\delta^{11/2})
+\mathcal{O}(\epsilon^{1})
,\nonumber\\
I_{11}&=
\frac{\mathcal{N}^2}{s^2}\left(
\frac{2 \delta ^{5/2}}{135 \pi ^3}
+\frac{2 \delta ^{7/2}}{105 \pi ^3}
+\frac{598 \delta ^{9/2}}{25515 \pi ^3}
\right)+\mathcal{O}(\delta^{11/2})
+\mathcal{O}(\epsilon^{1})
,\nonumber\\
I_{12}&=
\frac{\mathcal{N}^2}{s}\left(
\frac{4 \delta ^{7/2}}{945 \pi ^3}
+\frac{16 \delta ^{9/2}}{2835 \pi ^3}
+\frac{1964 \delta ^{11/2}}{280665 \pi ^3}
\right)+\mathcal{O}(\delta^{13/2})
+\mathcal{O}(\epsilon^{1})
,\nonumber\\
I_{13}&=
\frac{\mathcal{N}^2}{\epsilon s^3}\left(
-\frac{\delta ^{3/2}}{9 \pi ^3}
-\frac{2 \delta ^{5/2}}{15 \pi ^3}
-\frac{52 \delta ^{7/2}}{315 \pi ^3}
\right)+\mathcal{O}(\delta^{9/2})
+\mathcal{O}(\epsilon^{0})
,\nonumber\\
I_{14}&=
\frac{\mathcal{N}^2}{\epsilon s^3}\left(
-\frac{\delta ^{3/2}}{9 \pi ^3}
-\frac{2 \delta ^{5/2}}{45 \pi ^3}
-\frac{4 \delta ^{7/2}}{105 \pi ^3}
\right)+\mathcal{O}(\delta^{9/2})
+\mathcal{O}(\epsilon^{0})
,\nonumber\\
I_{15}&=
\frac{\mathcal{N}^2}{s^2}\left(
-\frac{2 \delta ^{5/2}}{45 \pi ^3}
-\frac{22 \delta ^{7/2}}{315 \pi ^3}
-\frac{694 \delta ^{9/2}}{8505 \pi ^3}
\right)+\mathcal{O}(\delta^{11/2})
+\mathcal{O}(\epsilon^{1})
,\nonumber\\
I_{16}&=
\frac{\mathcal{N}^2}{\epsilon s^3}\left(
\frac{\delta ^{3/2}}{9 \pi ^3}
+\frac{8 \delta ^{5/2}}{45 \pi ^3}
+\frac{218 \delta ^{7/2}}{945 \pi ^3}
\right)+\mathcal{O}(\delta^{9/2})
+\mathcal{O}(\epsilon^{0})
\,,
\end{align}
where
\begin{eqnarray}
  {\cal N} &=& x^\epsilon\left(\frac{\mu_r^2}{m_h^2}\right)^\epsilon
  \,.
               \label{eq::calN}
\end{eqnarray}

In principle, one can compute the series expansion of $I_i$ up to arbitrary
order in $\delta$ using the method described above.  However, the number of terms in
the integrand grows rapidly, and we stopped the computation at
$\mathcal{O}(\delta^{11/2})$. A more efficient approach to obtain
high-order terms in $\delta$ is based on differential equations
with respect to $\delta$.  It turns out that
for the boundary conditions required to solve the differential equations,
the leading-order term of each integral is sufficient. We can compute this
term without expanding in $\epsilon$ and obtain
\begin{align}
I_{1}&=\mathcal{N}^2 \left( \frac{e^{\gamma_E}}{4\pi} \right)^{2\epsilon}
\left[\frac{2^{2 (3 \epsilon-4)} \pi ^{2 \epsilon-\frac{5}{2}} s \delta ^{\frac{5}{2}-3 \epsilon} \Gamma (1-\epsilon)}{\Gamma \left(\frac{7}{2}-3 \epsilon\right)}
+\mathcal{O}(\delta^{7/2})
\right],\nonumber\\
I_{2}&=\mathcal{N}^2 \left( \frac{e^{\gamma_E}}{4\pi} \right)^{2\epsilon}
\left[-\frac{3\ 2^{2 (3 \epsilon-5)} \pi ^{2 \epsilon-\frac{5}{2}} s^2 \delta ^{\frac{5}{2}-3 \epsilon} \Gamma (1-\epsilon)}{\Gamma \left(\frac{7}{2}-3 \epsilon\right)}
+\mathcal{O}(\delta^{7/2})
\right],\nonumber\\
I_{3}&=\mathcal{N}^2 \left( \frac{e^{\gamma_E}}{4\pi} \right)^{2\epsilon}
\left[\frac{2^{6 (\epsilon-1)} \pi ^{2 \epsilon-\frac{5}{2}} \delta ^{\frac{5}{2}-3 \epsilon} \Gamma (1-\epsilon)}{\Gamma \left(\frac{7}{2}-3 \epsilon\right)}
+\mathcal{O}(\delta^{7/2})
\right],\nonumber\\
I_{4}&=\mathcal{N}^2 \left( \frac{e^{\gamma_E}}{4\pi} \right)^{2\epsilon}
\left[-\frac{2^{8 \epsilon-5} \pi ^{2 \epsilon-2} \delta ^{\frac{3}{2}-3 \epsilon} \Gamma (-2 \epsilon)}{s \Gamma \left(\frac{5}{2}-3 \epsilon\right) \Gamma \left(\frac{1}{2}-\epsilon\right)}
+\mathcal{O}(\delta^{5/2})
\right],\nonumber\\
I_{5}&=\mathcal{N}^2 \left( \frac{e^{\gamma_E}}{4\pi} \right)^{2\epsilon}
\left[\frac{2^{2 (3 \epsilon-2)} \pi ^{2 \epsilon-\frac{5}{2}} \delta ^{\frac{5}{2}-3 \epsilon} \Gamma (1-\epsilon)}{s \Gamma \left(\frac{7}{2}-3 \epsilon\right)}
+\mathcal{O}(\delta^{7/2})
\right],\nonumber\\
I_{6}&=\mathcal{N}^2 \left( \frac{e^{\gamma_E}}{4\pi} \right)^{2\epsilon}
\left[\frac{2^{8 \epsilon-5} \pi ^{2 \epsilon-2} \delta ^{\frac{3}{2}-3 \epsilon} \Gamma (-2 \epsilon)}{s \Gamma \left(\frac{5}{2}-3 \epsilon\right) \Gamma \left(\frac{1}{2}-\epsilon\right)}
+\mathcal{O}(\delta^{5/2})
\right],\nonumber\\
I_{7}&=\mathcal{N}^2 \left( \frac{e^{\gamma_E}}{4\pi} \right)^{2\epsilon}
\left[-\frac{2^{6 (\epsilon-1)} \pi ^{2 \epsilon-\frac{5}{2}} \delta ^{\frac{5}{2}-3 \epsilon} \Gamma (1-\epsilon)}{3 \Gamma \left(\frac{7}{2}-3 \epsilon\right)}
+\mathcal{O}(\delta^{7/2})
\right],\nonumber\\
I_{8}&=\mathcal{N}^2 \left( \frac{e^{\gamma_E}}{4\pi} \right)^{2\epsilon}
\left[\frac{2^{2 (3 \epsilon-2)} \pi ^{2 \epsilon-\frac{5}{2}} \delta ^{\frac{5}{2}-3 \epsilon} \Gamma (1-\epsilon)}{3 s \Gamma \left(\frac{7}{2}-3 \epsilon\right)}
+\mathcal{O}(\delta^{7/2})
\right],\nonumber\\
I_{9}&=\mathcal{N}^2 \left( \frac{e^{\gamma_E}}{4\pi} \right)^{2\epsilon}
\left[-\frac{2^{2 (3 \epsilon-2)} \pi ^{2 \epsilon-\frac{5}{2}} \delta ^{\frac{5}{2}-3 \epsilon} \Gamma (1-\epsilon)}{3 s \Gamma \left(\frac{7}{2}-3 \epsilon\right)}
+\mathcal{O}(\delta^{7/2})
\right],\nonumber\\
I_{10}&=\mathcal{N}^2 \left( \frac{e^{\gamma_E}}{4\pi} \right)^{2\epsilon}
\left[-\frac{2^{2 (3 \epsilon-1)} \pi ^{2 \epsilon-\frac{5}{2}} \delta ^{\frac{5}{2}-3 \epsilon} \Gamma (1-\epsilon)}{3 s^2 \Gamma \left(\frac{7}{2}-3 \epsilon\right)}
+\mathcal{O}(\delta^{7/2})
\right],\nonumber\\
I_{11}&=\mathcal{N}^2 \left( \frac{e^{\gamma_E}}{4\pi} \right)^{2\epsilon}
\left[\frac{2^{2 (3 \epsilon-1)} \pi ^{2 \epsilon-\frac{5}{2}} \delta ^{\frac{5}{2}-3 \epsilon} \Gamma (1-\epsilon)}{9 s^2 \Gamma \left(\frac{7}{2}-3 \epsilon\right)}
+\mathcal{O}(\delta^{7/2})
\right],\nonumber\\
I_{12}&=\mathcal{N}^2 \left( \frac{e^{\gamma_E}}{4\pi} \right)^{2\epsilon}
\left[\frac{2^{2 (4 \epsilon-2)} \pi ^{2 \epsilon-2} \delta ^{\frac{7}{2}-3 \epsilon} \Gamma (3-2 \epsilon)}{9 s \Gamma \left(\frac{9}{2}-3 \epsilon\right) \Gamma \left(\frac{3}{2}-\epsilon\right)}
+\mathcal{O}(\delta^{9/2})
\right],\nonumber\\
I_{13}&=\mathcal{N}^2 \left( \frac{e^{\gamma_E}}{4\pi} \right)^{2\epsilon}
\left[\frac{2^{8 \epsilon-1} \pi ^{2 \epsilon-2} \delta ^{\frac{3}{2}-3 \epsilon} \Gamma (-2 \epsilon)}{3 s^3 \Gamma \left(\frac{5}{2}-3 \epsilon\right) \Gamma \left(\frac{1}{2}-\epsilon\right)}
+\mathcal{O}(\delta^{5/2})
\right],\nonumber\\
I_{14}&=\mathcal{N}^2 \left( \frac{e^{\gamma_E}}{4\pi} \right)^{2\epsilon}
\left[\frac{2^{8 \epsilon-1} \pi ^{2 \epsilon-2} \delta ^{\frac{3}{2}-3 \epsilon} \Gamma (-2 \epsilon)}{3 s^3 \Gamma \left(\frac{5}{2}-3 \epsilon\right) \Gamma \left(\frac{1}{2}-\epsilon\right)}
+\mathcal{O}(\delta^{5/2})
\right],\nonumber\\
I_{15}&=\mathcal{N}^2 \left( \frac{e^{\gamma_E}}{4\pi} \right)^{2\epsilon}
\left[-\frac{2^{2 (3 \epsilon-1)} \pi ^{2 \epsilon-\frac{5}{2}} \delta ^{\frac{5}{2}-3 \epsilon} \Gamma (1-\epsilon)}{3 s^2 \Gamma \left(\frac{7}{2}-3 \epsilon\right)}
+\mathcal{O}(\delta^{7/2})
\right],\nonumber\\
I_{16}&=\mathcal{N}^2 \left( \frac{e^{\gamma_E}}{4\pi} \right)^{2\epsilon}
\left[-\frac{2^{8 \epsilon-1} \pi ^{2 \epsilon-2} \delta ^{\frac{3}{2}-3 \epsilon} \Gamma (-2 \epsilon)}{3 s^3 \Gamma \left(\frac{5}{2}-3 \epsilon\right) \Gamma \left(\frac{1}{2}-\epsilon\right)}
+\mathcal{O}(\delta^{5/2})
\right]\,.
\end{align}
We managed to obtain all master integrals up to $\mathcal{O}(\delta^{219/2})$
within a few hours of CPU time which is sufficient for all practical
purposes. Note, however, that with this approach the computation of further
orders in the expansion is not difficult; we have produced expressions for the
expansion up to $\mathcal{O}(\delta^{1019/2})$.

It turns out that all of the master integrals have non-integer powers of
$\delta$ after expansion.  An overall factor $\delta^{1/2}$ comes from the
measure of the $k$ integration in Eq.~\eqref{eq:app:int}.  Terms with odd
powers of $k$ in the integrand are candidates for terms with integer power of
$\delta$. However, these vanish after the angular integration.  Since the
differential equation relates terms whose powers of $\delta$ differ by
integers, we have to confirm the absence of integer powers in $\delta$ by an
explicit calculation of the first few terms, as we have shown above.


\section{\label{app::MI_exact}Exact master integrals}

\subsection{Two-loop master integrals}

In order to obtain exact results for the master integrals we first transform
the differential equation into a canonical
form~\cite{Henn:2013pwa}. Afterwards we perform the integrations order by
order in $\epsilon$ and express the analytic results in terms of Goncharov
polylogarithms~\cite{Goncharov:1998kja} which can be evaluated numerically
using {\tt GiNaC}~\cite{Bauer:2000cp,Vollinga:2004sn}.

To obtain a canoncial form, we apply Lee's algorithm~\cite{Lee:2014ioa} as implemented in the program
{\tt Epsilon}~\cite{Prausa:2017ltv} to bring the $16\times 16$ system to normal
Fuchsian form (i.e. $\epsilon$ does not yet factorize) and observe that the
differential equations contain poles at
$x = \{0,1/4,1,r_1=\exp(i\pi/3),r_2=\exp(-i\pi/3),-1/3\}$. The first and
second poles correspond to the limits $s\rightarrow\infty$ and
$s\rightarrow 4 m_H^2$, respectively.  The remaining four poles are only
present in sectors with a third, uncut Higgs propagator, i.e. for integrals
$I_7$ to $I_{16}$.  Note that $x = 1$ corresponds to the threshold for single
Higgs production.

Let us first have a closer look at the sub-matrix which corresponds to $I_1$
and $I_2$. After Fuchsification the matrix residue at $x=1/4$ for the subsystem
reads
\begin{align}
\mathcal{M}^{(1,2)}_{\frac{1}{4}} =
  \begin{pmatrix}
    0 & 0 \\
    0 & \frac{5}{2}-3\epsilon
  \end{pmatrix}
  \,,
\end{align}
which indicates that one of the eigenvalues is half-integer for
$\epsilon\rightarrow 0$.  This implies the appearance of the square root
$\sqrt{1-4x}$ in the alphabet, which is due to the two-particle branch cut.
We rationalize this root by introducing a new variable $y$, defined as
\begin{align}
  y = \frac{\sqrt{1-4x}-1}{\sqrt{1-4x}+1},~-1 < y < 0\,.
  \label{eq::y}
\end{align}
After rationalization we managed to find a canonical basis for 
the first 14 master integrals.

The homogeneous part of the differential equation for $I_{15}$ contains another
residue with half-integer eigenvalue, namely $1/(y^4-3y^3+5y^2-3y+1)$.
After re-scaling $I_{15}$ we are able to bring the
differential equations into a form in which we can also factor out $\epsilon$
and the homogeneous parts of $I_{15}$ and $I_{16}$ only contain single poles
in $y$. The root $\sqrt{1-6y+7y^2-6y^3+y^4}$ appears now in the inhomogeneous
contributions. Note, however, that only the leading terms in
$\epsilon$ of these two integrals are needed, which means that we do not have
to iteratively integrate over the square root.

To simplify the integration of the differential equations and the manipulation
of the amplitudes, we do not partial fraction the polynomials
\begin{align}
  P_2 &= y^2-y+1~,\nonumber\\
  P_4 &= y^4-3y^3+5y^2-3y+1
\end{align}
which appear in the denominators
of the differential equations. Rather, we use the following 
integration kernels
\begin{align}
  f(0;y)=\frac{1}{y},~f(1;y) = \frac{1}{y-1},~f(-1;y) = \frac{1}{y+1},\nonumber\\
  f(r^{(n)};y) = \frac{\partial^n_y P_2}{P_2},~f(s^{(k)};y) =
  \frac{\partial^{k}_y
  P_4}{P_4},
\label{eq::kernels}
\end{align}
in our final expressions, where $n=1,2$ and $k=1,\ldots,4$.
Note that for the numerical evaluation we can partial
fraction the quadratic and quartic kernels and rewrite the integrals
as a sum of Goncharov polylogarithms, i.e.
\begin{align}
G(...,r^{(n)},...;y) &= \sum_{i=1}^2c_i^{(n)}G(...,r_i,...;y)\,,\nonumber\\
G(...,s^{(k)},...;y) &= \sum_{i=1}^4c_i^{(k)}G(...,s_i,...;y)\,,
\label{eq::partfrac}
\end{align}
where the $r_i$ are the roots of $P_2$ and the $s_i$ are the roots of
$P_4$.  While all iterated integrals over the kernels in
Eq.~\eqref{eq::kernels} are real-valued for $-1 \le y \le 0$, the individual
Goncharov polylogarithms on the r.h.s. of Eq.~\eqref{eq::partfrac} are
not.

We expand all master integrals in $\epsilon$ up to the order needed
for the finite NNLO part, which means that some master integrals are
expanded up to the $\epsilon^2$ and for others only the $1/\epsilon$
pole is required.  Let us mention that the
$\mathcal{O}\left(\epsilon^2\right)$ terms of the master integrals
$I_9$ and $I_{12}$ contain Goncharov polylogarithms up to weight~4.
However, only the difference $I_9-I_{12}$, which only contains
weight-3 Goncharov polylogarithms, is needed up to
$\mathcal{O}\left(\epsilon^2\right)$. The sum $I_9+I_{12}$ is only
needed up to the linear $\epsilon$ term and contains at most weight-3
terms.  Therefore the cross section can be expressed in terms of Goncharov
polylogarithms up to weight~3.

In the ancillary file~\cite{progdata} we provide analytic expressions
for all masters integrals, both expanded in $\delta$ up to order
$\delta^{219/2}$ and expressed in terms of Goncharov
polylogarithms. For the latter we do not give separate expressions for
$I_9$ and $I_{12}$ but for the combinations $I_9-I_{12}$ and $I_9+I_{12}$.

\begin{figure}[t]
  \begin{center}
    \begin{tabular}{cc}
      \includegraphics[width=.5\textwidth]{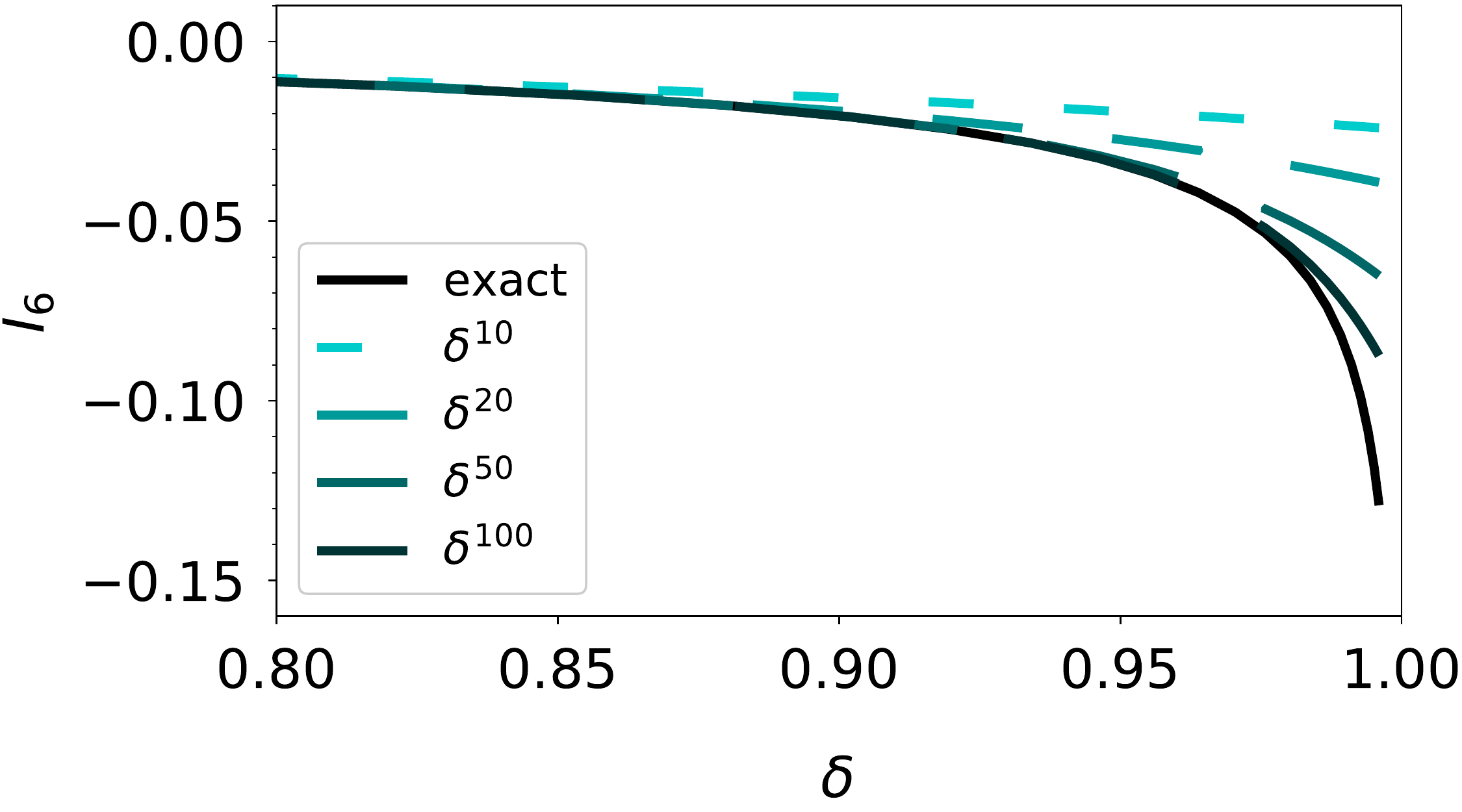} &
      \includegraphics[width=.5\textwidth]{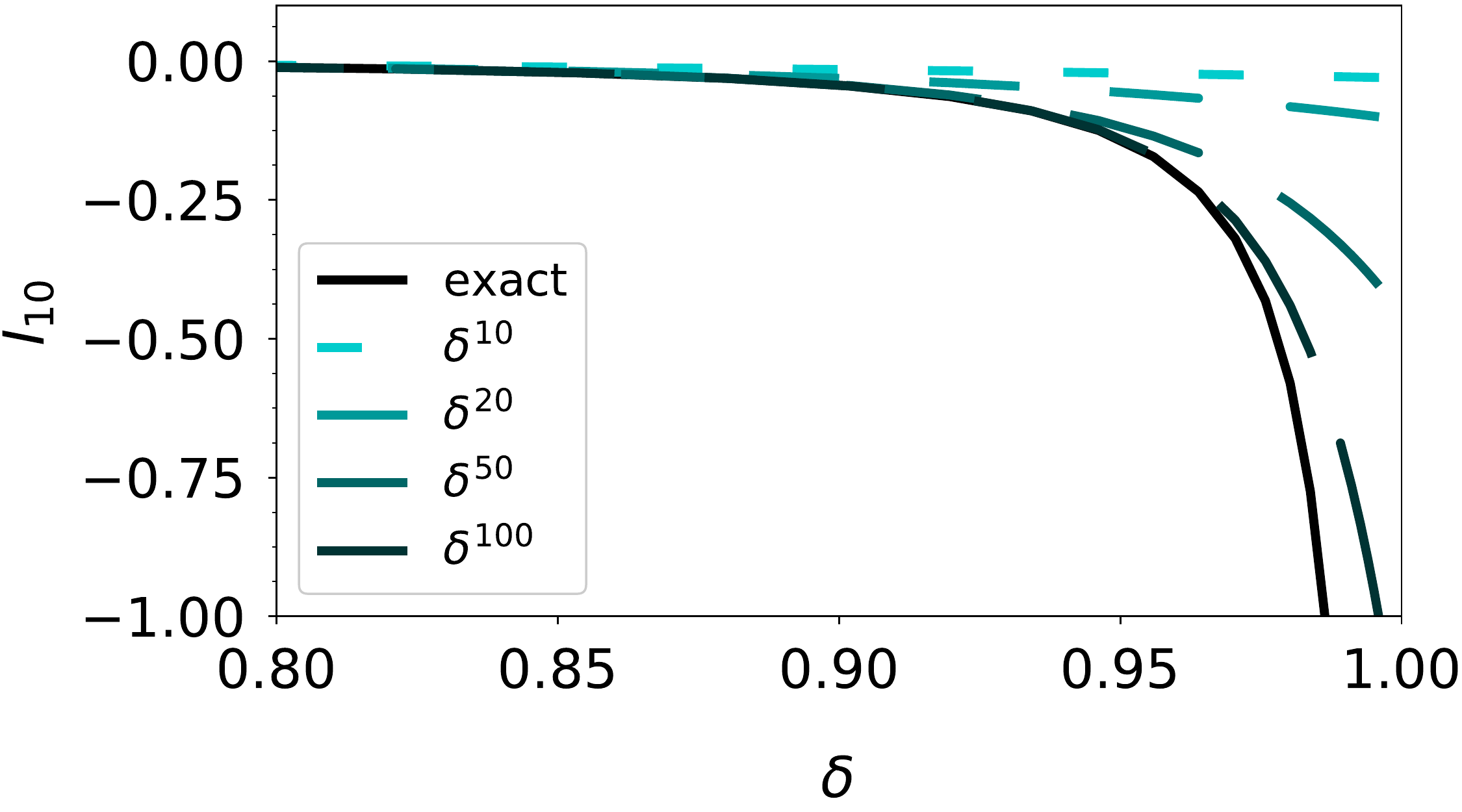}
    \end{tabular}
  \end{center}
  \caption{\label{fig::mi}The finite parts of the master integrals $I_6$ and $I_{10}$
    as a function of $\delta$. The exact and expanded results are shown as
    solid and dashed lines, respectively. The plot legends show the expansion depths
    included in the plots.}
\end{figure}

We are now in a position to compare the exact results for the master integrals
with the $\delta$-expanded expressions. In Fig.~\ref{fig::mi} we show, for two
typical examples, the exact result (solid curve) and results expanded up to
various orders in $\delta$ (dashed curves).  We plot the expressions as a
function of $\delta$ but suppress the threshold region where very good
agreement is found. The expressions expanded up to $\delta^{10}$ start to
deviate from the exact curve above $\delta\approx 0.8$. Agreement up to
$\delta\approx 0.9$ is observed if 20 expansion terms in $\delta$ are included
and for 100 terms we reach $\delta\approx 0.97$.  Note that $\delta=0.9$
corresponds to $\sqrt{s}\approx 800$~GeV where the parton distribution
functions are already quite small. For many applications it is therefore
sufficient to work with the $\delta$-expanded expressions.

During preparation of this manuscript, the \texttt{Mathematica} package
\texttt{PolyLogTools}~\cite{Duhr:2019tlz} was made available. We were able to
use it to expand the exact expressions for our master integrals in $\delta$ to
order $\delta^{11/2}$, and found full agreement with our expansions of
Appendix~\ref{app::MI_delta}.

\subsection{One-loop master integrals}

The two one-loop master integrals (see Fig.~\ref{fig::masters1l}) have
been computed in Ref.~\cite{Grigo:2013rya} as an expansion in $\delta$. We
have solved the system of differential equations using the same approach as at
two loops (see above) and obtained the following results which are
exact in $y$ (see Eq.~(\ref{eq::y})):
\begin{eqnarray}
  J_1 
  &=& 
      \mathcal{N}\left(\frac{e^{\gamma_\mathrm{E}}}{4\pi}\right)^\epsilon\left(\frac{y+1}{8\pi(1-y)}\right)
      \Bigg[1 + 2\epsilon\left(1 - G\left(-1;y\right) + G\left(1;y\right)\right)\nonumber\\&&
      + 4\epsilon^2\left(1-\frac{3\zeta_2}{8} - G\left(-1;y\right) + G\left(1;y\right) + \frac{1}{2}\left(G\left(-1;y\right) - G\left(1;y\right)\right)^2\right) + \mathcal{O}\left(\epsilon^3\right) \Bigg]
      \,,
  \nonumber\\
  J_2 
  &=&
      \mathcal{N}\left(\frac{e^{\gamma_\mathrm{E}}}{4\pi}\right)^\epsilon\left(\frac{1}{4\pi}\right)
      \Bigg[-G\left(0;-y\right) + \epsilon\Big(\zeta_2 + 2G\left(0,-1;y\right) + 2G\left(0,1;y\right)\nonumber\\&& - 4G\left(1;y\right)G\left(0;-y\right) + G\left(0;-y\right)^2\Big)
      + \epsilon^2\Big(\zeta_3 + 4\zeta_2G\left(1;y\right) - \frac{\zeta_2}{2}G\left(0;-y\right)\nonumber\\&& 
      - 4 \Big(G\left(0,-1,1;y\right) + G\left(0,1,-1;y\right) + G\left(0,-1,-1;y\right) + G\left(0,1,1;y\right)\nonumber\\&&  
      + G\left(0,-1,-1;y\right) + G\left(0,0,1;y\right) + G\left(0,0,-1;y\right)\Big)\Big) - 8G\left(1;y\right)^2G\left(0;-y\right)\nonumber\\&&
      + 8G\left(1;y\right)\Big(G\left(0,1;y\right)+G\left(0,-1;y\right)\Big) + 4G\left(1;y\right)G\left(0;-y\right)^2 - \frac{2}{3}G\left(0;-y\right)^3\nonumber\\&& + \mathcal{O}\left(\epsilon^3\right) \Bigg]
      \,,
\end{eqnarray}
where ${\cal N}$ is given in Eq.~(\ref{eq::calN}).




\end{document}